# White paper on cybersecurity in the healthcare sector. The HEIR solution.

Lampropoulos K. (editor), Zarras A., Lakka E., Barmpaki P., Drakonakis K., Athanatos M.[1], Herve D.[2], Alexopoulos A., Sotiropoulos A.[3], Tsakirakis G., Dimakopoulos N.[4], Tsolovos D., Pocs M.[5], Smyrlis M. (editor), Basdekis I., Spanoudakis G.[6], Mihaila O., Prelipcean B.[7], Salant E.[8], Athanassopoulos S., Papachristou P.[9], Ladakis I.[10], Chang J.[11], Floros E. (editor), Smyrlis K.[12], Besters R., Årsand E., Randine P.[13], Løvaas K.F., Cooper J.[14], Ilie I., Gabriel D.[15], Khabbaz M.[16]

IDRYMA TECHNOLOGIAS KAI EREVNAS[1]
INSTITUT MINES-TELECOM[2]
AEGIS IT RESEARCH GMBH[3]
IOTAM INTERNET OF THINGS APPLICATIONS AND MULTI LAYER DEVELOPMENT LTD[4]
STELAR SECURITY TECHNOLOGY LAW RESEARCH UG[5]
SPHYNX TECHNOLOGY SOLUTIONS AG[6]
BITDEFENDER SRL[7]
IBM ISRAEL - SCIENCE AND TECHNOLOGY LTD[8]
HYGEIA[9]
WELLICS LTD[10]
CROYDON HEALTH SERVICES NATIONAL HEALTH SERVICE TRUST[11]
PANEPISTIMIAKO GENIKO NOSOKOMEIO IRAKLEIOU[12]
UNIVERSITETSSYKEHUSET NORD- NORGE HF[13]
HARALDSPLASS DIAKONALE SYKEHUS AS[14]
SIEMENS SRL[15]
TU DELFT[16]

**Abstract**: The healthcare sector is increasingly vulnerable to cyberattacks due to its growing digitalization. Patient data, including medical records and financial information, are at risk, potentially leading to identity theft and patient safety concerns. The European Union and other organizations identify key areas for healthcare system improvement, yet the industry still grapples with inadequate security practices. In response, the HEIR project offers a comprehensive cybersecurity approach, promoting security features from various regulatory frameworks and introducing tools such as the Secure Healthcare Framework and Risk Assessment for Medical Applications (RAMA). These measures aim to enhance digital health security and protect sensitive patient data while facilitating secure data access and privacy-aware techniques. In a rapidly evolving threat landscape, HEIR presents a promising framework for healthcare cybersecurity.

Contents

# 1  Introduction

The healthcare sector has increasingly become a prime target for cyberattacks. According to a report by ENISA from 2023 [1], the healthcare industry is one of the most vulnerable to cybersecurity incidents. In a healthcare environment that's becoming increasingly digital, crucial data like medical records, personal identifiers, and financial information are stored and transmitted electronically. Safeguarding this data is of utmost importance because cyberattacks can lead to identity theft, fraud, or ransomware events that could jeopardize patient safety and shake trust in the healthcare system. Furthermore, as telemedicine and similar trends continue to grow, weaknesses in cybersecurity pose the potential for life-threatening interruptions in healthcare services. Healthcare providers can ensure patient privacy, preserve the integrity of medical data, and guarantee the delivery of high-quality, secure healthcare services by making substantial investments in robust cybersecurity measures. The majority of EU countries allocate substantial resources to procure the latest e-health tools and applications, with the goal of providing effective and efficient healthcare services to their citizens [2]. These services are designed to facilitate the seamless exchange of health information, ultimately enhancing the interaction between healthcare professionals and their patients.

Regrettably, the healthcare industry has not only emerged as a primary target for cybercriminals but also witnessed the costliest data breaches over the past 13 years [3]. This alarming trend has been exacerbated by the COVID-19 pandemic, as highlighted in Muthuppalaniappan's report [4]. There are two fundamental factors contributing to this troubling phenomenon. First, healthcare facilities possess highly valuable assets. Furthermore, a patient's aggregated data can be likened to a valuable goldmine, encompassing a comprehensive biography that includes essential details, health patterns, family history, and even financial information. Secondly, healthcare organizations are susceptible to easy compromise. Notably, Kumar [5] underscores that medical data holds greater value than financial data since healthcare records can be exploited long after the initial security breach that exposed them. This underscores the critical importance of bolstering cybersecurity measures in the healthcare sector.

Recognizing the challenges faced by healthcare systems, three key areas for improvement have been recognized by the EU and other organizations and institutes:

1. Ensuring citizens have secure access to electronic health records that can be shared across borders.
2. Supporting data infrastructure to advance research, disease prevention, and personalized health and care.
3. Facilitating feedback and interaction between patients and healthcare providers, enhancing disease prevention, and empowering individuals to take responsibility for their health.

Despite the significant strides made in the healthcare sector to harness digital advancements for enhancing patient outcomes and experiences, there remains a prevalent reliance on poor security practices [6]. As pointed out by Coventry et al. [7], crucial issues in the healthcare sector include inadequate computer and user account security, challenges with remote access and home working, and a lack of robust encryption measures. These issues, compounded by outdated risk assessment techniques and a deficit in security awareness, have rendered the healthcare sector highly susceptible to substantial data breach costs, as highlighted in. Additionally, it has become the sector most besieged by cyber-attacks, both in terms of volume (comprising 69%) and complexity (constituting 67%). [8]

Considering these challenges, the HEIR project worked on a comprehensive approach to cybersecurity which requires all cyber systems must incorporate security features stemming from multiple regulatory frameworks such as GDPR, ENISA, NIS etc. Targeting to enhance digital health security across the EU, the HEIR project built a framework to reduce market access limitations, conflicting requirements, administrative burdens etc., and eventually establish a broad European network dedicated to promoting cybersecurity best practices across all regulatory frameworks. More specifically, HEIR developed a Secure Healthcare Framework that facilitates intelligent threat identification and hunting services, ML-based anomaly detections, as well as privacy-aware techniques, leading to the delivery of a secure and holistic solution. Moreover, HEIR introduced the Risk Assessment for Medical Applications (RAMA) score. This score acts as a risk assessment and estimates the attack surface and resilience of medical applications by incorporating several critical issues. The proposed solution is further separated into the Local and Global RAMA Scores to address the need to calculate it locally (within a single healthcare organisation) and globally (within all the assessed healthcare organisations). The former incorporates several critical issues reported by a centralised client and estimates the attack surface and resilience of the underlying medical applications per healthcare organization, whereas the latter serves as a global benchmark against which local RAMA scores will be compared.

More to that, HEIR introduced advanced visualisations allowing end users to examine the output of the aforementioned tools. More specifically, HEIR's 1$^{st}$ Layer GUI and the Forensics Visualisation Toolkit (FVT) are two tools that provide insights pertinent to the hospital's local environment, whereas HEIR's Observatory provides, anonymized, aggregated data targeting external stakeholders.

Lastly, HEIR introduced the Privacy Aware Framework [9]. HEIR's PAF is built on top of the Open-Source project, Fybrik. Extending Fybrik with a custom FybrikModule designed around HEIR requirements, we created a prototype application that allows third-party researchers to execute ad-hoc queries on a hospital's FHIR server, subject to policy-defined data constraints. Leveraging Fybrik's Policy Manager connector, we use Open Policy Agent as our data governance engine and the Rego language to define HEIR's fine-grained, powerful policy rules for the protection of FHIR data. These rules can take into account virtually any type of access requirements, such as the geographical locations of the data requester and store, and the intended use of the data.

This allows third parties to obtain access to the hospital's FHIR records by working with the hospital's Data Governance Officer to determine which data will be provided, and the context under which data will be made available.

## 1.1 Identified challenges towards the development of secure and trustworthy healthcare systems and services.

This section outlines the challenges that have been identified when assessing the implementation of HEIR in the context of the security of healthcare sector. In particular, the challenges are linked to the core of the solution offered.

*Challenge 1 – Healthcare data breaches:* Healthcare data breaches are a growing threat to the healthcare industry, causing data loss and monetary theft, and attacks on medical devices and infrastructure. Hospital data security breaches can cost a single hospital a substantial financial fine, litigation, and a damaged reputation. Meanwhile, the healthcare industry lags other industries in securing its data. In response, healthcare organizations must invest considerable capital and effort in protecting their systems.

*Challenge 2 – Vulnerable medical devices:* As billions of medical devices will be imported into the healthcare domain, the impact is expected to be significant. Healthcare providers have a unique opportunity to use the data from these devices to improve patient outcomes. Still, they need to find ways to get insight into so much data so it can be actionable. Medical IoT will extend the connectivity and transmission of health data from the patient to the physician regularly or immediately and continuously in an emergency. The Medical IoT will make medicine participatory, personalized, predictive and preventive (P4 Medicine). Despite the evident material gains due to the increased digital connectivity, these technological advancements in the healthcare domain often come with security risks due to its novelty and complexity. As the number of connected devices and cloud networks increases, the attack surface for data breaches or ransomware becomes greater than ever before, and the need for innovative technologies comprising a wide range of fields becomes inevitable to counter such attacks.

*Challenge 3 – Privacy-sensitive data:* Despite the wide range of tools and services already available to facilitate the operations in hospitals and medical centers, that domain still lacks innovative, secure execution environments based on novel tools and services that can establish secure digital collaboration, especially under the challenges introduced by continuously updated GDPR requirements and the growing exploitation of IoT-based medical devices and wearables. Ethical considerations are playing catch-up: anonymized data sets could be harvested and analyzed without the patient's consent to the use of their data. There is a need for the patient to say how their electronic data can be used and by whom.

*Challenge 4 – Legacy systems:* Field-specific challenges (many healthcare systems are in existence within various localities; systems do not communicate efficiently, leading to patients transporting their medical records in paper format between hospitals; records not always available on time and can be easily lost and left open for public scrutiny) impose yet another need for such validated systems in real-life environments, addressing issues of Information Governance and privacy concerns.

*Challenge 5 – User awareness:* Security policy, governance and end-user awareness need to extend across all processes and levels of healthcare environments as complex systems become more and more interconnected. Moreover, the lack of security awareness across the environment and fragmented security solutions that don't necessarily work from one system to another (e.g., applying IT resources such as invasive penetration testing and network mapping tools to different departments of medical centers), are major hurdles and roadblocks to exploiting advanced technologies' full potential in interconnected healthcare environments.

*Challenge 6 – Trust increase:* Cyber-crime and attacks against critical infrastructures affect the economy and business growth in multiple ways. Achieving a high degree of trust in EU digital networks, products and services requires multidisciplinary research on longer-term security challenges complemented by non-technical aspects of cybersecurity and digital privacy such as business viability and business alliances and collaborations.

## 2 State of the art.

This section provides a comprehensive analysis of the state-of-the-art, highlighting the advances of HEIR, both for the HEIR platform as a whole and for each specific scientific and technological domain of interest to HEIR.

### 2.1 Security and privacy assessment

Healthcare environments such as hospitals are increasingly relying on connected devices, large and small. Such environments' complexity makes risk management extremely difficult, given the large attack surface [10]. Therefore, cyber-risk assessment is becoming a critical part of running a hospital's IT network, as indicated in the report from the Health Care Industry Cybersecurity Task Force [11]. This report's first imperative cites this methodological aspect as the top priority for cybersecurity in healthcare. The establishment of such a reference cybersecurity framework and methodology for healthcare is specifically mentioned as a recommendation of the report, as well as the establishment of scalable best practices for governance.

While there are several general-purpose cybersecurity methodologies available, the most well-known being the Cybersecurity Framework proposed by the *National Institute of Standards and Technology (NIST)*, there is no such document that helps to manage the many peculiar and often conflicting requirements of the healthcare environments. Work carried out in HEIR related to the RAMA score, and the observatory is particularly relevant to support up to date and quantified cybersecurity and privacy risk assessment.

## 2.2 Security and privacy preservation

One of the foremost issues of medical environments is the protection of data, be it medical or administrative, in an environment where expressing security policies is difficult. The medical world is increasingly relying on its data for diagnosis and treatment, therefore on sensors and devices to create this data and communications channels to collect, store, and exchange it. Medical activity is likely to become mostly analytics-driven, taking advantage of the volume of data collected by large and small devices. This creates both new opportunities and new threats. The Health Care Industry Cybersecurity Task Force report [11] provides the need for protection of data is further reinforced by the entry into law of the GDPR [12]. HEIR was based on standards, specifically Health Level 7 (HL7), to enhance its solutions' applicability. The most prevalent system for describing EPR is the Health Level Seven (HL7) Clinical Document Architecture (CDA) [13] [14]. Another critical aspect is the sharing of data through cloud computing and storage solutions. The latter has become very common among enterprises in general and among hospitals and healthcare facilities in particular. One of the main benefits is the simplification of the information sharing process among multiple organizations or departments, which is a fundamental requirement in the healthcare world: the patients meet many practitioners (doctors, nurses, etc.) and organizations (hospitals, laboratories, nursing homes, etc.) over a long period of time: efficient and secure information sharing strategies among these stakeholders can have a huge impact when the lives of patients are on the line. However, storing sensitive health records in the cloud, whilst enabling increased availability, exposes the security and the privacy of these records to the risk of being violated [15] [16]. Recent developments in cloud architectures have originated new models of online storage clouds based on data dispersal algorithms [17] [18] [19]. Existing solutions have been explicitly applied in the healthcare world to address the well-known issues of privacy and confidentiality that arise when patients' data are transferred to remote cloud storage services [20] [21] [22] [23] [24]. Ensuring confidentiality in this context is crucial: only legitimate users should access any part of the information they distribute among storage nodes. The key idea behind all such solutions is that the data is divided into several distributed pieces among remote and independent storage nodes. Formal analysis techniques have been employed to assess their degree of confidentiality against honest-but curious cloud storage providers and external attackers [25].

## 2.3 Electronic medical devices security and trust

Electronic medical devices (EMDs) offer a plethora of possibilities in the healthcare domain, aiming to increase the ability of healthcare providers to treat patients and improve healthcare overall. They provide services for better patient monitoring, early and more precise diagnosis, online medical treatment, disease prevention, automated control, and central reporting and monitoring of data. These services involve access to personal medical data. These services can also be offered across borders, giving citizens the feeling of security in this respect. However, for all stakeholders to fully benefit from and trust electronic medical devices, they must be appropriately designed, implemented cost-effectively, and provide an acceptable level of security and privacy. The cybersecurity of medical devices is a complex and challenging ecosystem, which gradually has become a major concern to healthcare organizations, device manufacturers, and patients (Figure 1).

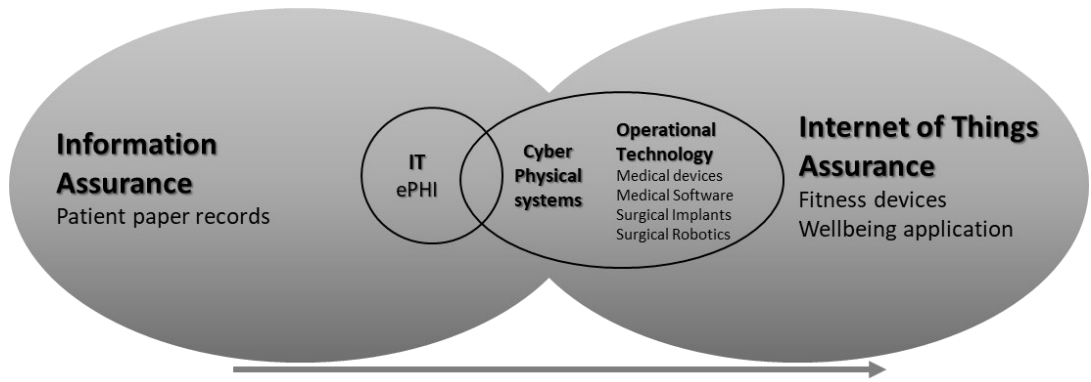

*Figure 1: The healthcare cybersecurity landscape is changing rapidly.*

EMDs are increasingly connected to the Internet, hospital networks, and other medical or smart devices to provide their features, which increases the risk of potential cybersecurity threats. The computer technology and software, as well as the convergence of networking, has allowed the integration of healthcare systems with Information Technology (IT) through remote accessibility, facilitating the revolution of cloud-based services, and the usage of "big" data analytics. Medical devices currently integrated with an increasingly digital healthcare infrastructure are vulnerable to the same security threats, as any other IT component [26], potentially impacting the safety and effectiveness of the device. Threats and vulnerabilities cannot be eliminated; therefore, reducing cybersecurity risks is especially challenging. Over the years, commercial medical solutions have complied with the minimal *Food and Drug Administration (FDA)* regulations on the medical devices and their data and have included secure communications. However, due to the long lifecycle of such devices, updating the equipment to purchase new hardware components or even performing software updates is not easy. The security of medical devices is an aspect that must be separately addressed in all involved technologies i.e., the medical devices themselves or other devices that are used, the different networks-wireless connections, and the healthcare delivery platforms [27] [28].This can be further complicated, if additional connected devices such as smart phones and tablets provide the healthcare service (Figure 2).

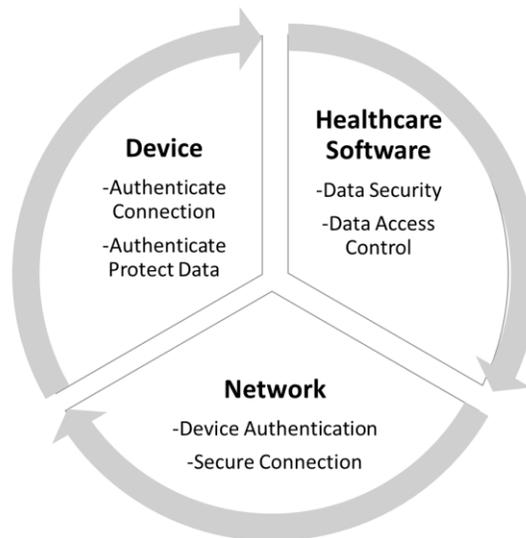

*Figure 2: The security of medical devices.*

Based on an ENISA study from 2016 [29], we can identify four principles for secure products, services and processes:

- Security by design – the product, service or process has been conceived, designed and implemented to ensure the key security properties are maintained: availability, confidentiality, integrity and accountability;

- Security by default – the product, service or process is supplied with the confirmed capability to support these security properties at installation;

- Security throughout the lifecycle – security should be maintained from initial deployment through maintenance to decommissioning;

- Verifiable security – each of the above principles should be verifiable.

Based on these principles, the main cybersecurity and privacy research challenges related to HEIR are described below. Key points to be considered in wireless medical devices is the fact that security and cryptography functions must have very small power consumption fingerprint, key agreement and authentication, authorization must be done without patient involvement and information must not leak from the device through some side channel.

### 2.4 Interoperable and scalable security management in heterogeneous ecosystems

The definition of security management policies to deal with heterogeneity and interoperability across domains, systems and networks, introduces several challenges related to the employed security models, the language and the level of abstraction required to govern the systems. This issue is exacerbated in healthcare deployments which are comprised of heterogenous disparate data sources and networks protocols/systems. Article 20(1) of the GDPR states that a data subject has the right to "receive the personal data concerning him or her, which he or she has provided to a controller, in a structured, commonly used and machine-readable format", and the right to "transmit those data to another controller". The GDPR article implies the need to have interoperability between different information systems. Popular international standards for interoperability are openEHR24/05/2019 21:54:00 [30], HL7 [31] and HL7 FHIR [32].

Medical devices (i.e., ECG recorder, blood pressure recorder, clinical laboratory systems for blood samples, etc.) generate different data and need to interoperate with the electronic health record implemented in the healthcare organization. Therefore, structural and semantic interoperability is an inherent part of the data generation. In addition, foundational interoperability lets the data transmitted by one healthcare information system to be received by another.

### 2.5 Reliable and privacy-preserving identity management and user authentication

Identity management systems require new security and privacy mechanisms that can holistically manage user's/object's privacy, ID-proofing techniques based on multiple biometrics, strong authentication, usage of breeder documents (e.g., eID, ePassports), while ensuring privacy-by-default, unlikability, anonymity, federation support, non-reputation and self-sovereign identification management. The challenge is to manage properly all these features for mobile, online or physical/face-to-face scenarios, while maintaining usability and compliance with regulation e.g., GDPR [12] and eIDAS [33]. This will ultimately lead to reduction in identity-theft and related cybercrimes.

Key generation and key agreement in *Body Area Network (BAN)* for wireless medical devices have been proposed that relies on using biometrics, or physiological values (PVs) [34] like Electrocardiograms (ECGs) [35] which relies on time interval between heartbeats, or interpulse interval (IPI) randomness.

Biometrics are nowadays a popular form of user authentication due to their ease of use, robustness and uniqueness compared to traditional knowledge-based systems, such as PINs and passwords. This has been seen especially on smartphones, where the use of fingerprints and face authentication to unlock the device is becoming more prevalent. Moreover, the wide availability of mobile sensors allows for the deployment of near frictionless multi-modal user authentication systems. Behavioral biometrics [36] are a particular kind of authentication factor that verifies the identity of users by the way they behave. They operate in the background in a continuous manner while the user interacts with an application. Typical examples of behavioral biometrics are keyboard dynamics [37] and mouse movements [38], and voice biometrics [39]. Sensor based gait recognition is also explored as a solution for unobtrusive user authentication [40] [41] [42] [43] with a wide applicability in wearable devices [44], [45]. By enabling continuous user authentication, gait authentication is a natural candidate for multi-modal settings, i.e., combining different types of biometric authentication factors. In this way, we can not only improve the accuracy of the user authentication system, but also strengthen the system against forging and spoofing attacks, while offering a user-friendly experience.

However, as soft biometrics like age, gender or race are linked to physiological or behavioral traits of the user, misuse of biometric templates may lead to severe privacy leakages for the user [46]. Previous work [47] [48] [49] has already shown the presence of sensitive data in biometric traces, including medical conditions and soft biometrics. In the case of gait-based user authentication, Garofalo et al. [50] demonstrated the feasibility of age and gender estimation from gait traces in the frame of the OU-ISIR Wearable Sensor-based Gait Challenge: *Age and Gender (GAG)* competition. To overcome these challenges, researchers have developed schemes which can detect the wearer of a wearable device through their physiological signals [51] [34] like ECGs [35], [52], relying on time between heartbeats, or *Interpulse Interval (IPI)* randomness, or bioelectrical impedance signals [53].

Privacy as Control can be implemented through *Privacy Enhancing Technologies (PET)*, ensuring selective and minimal disclosure of credentials and personal attributes, e.g., Anonymous Credential Systems [54], such as Idemix [55], which employs *Zero Knowledge Proofs (ZKPs)* reveal the minimal amount of information to the verifier (usually a service provider), even without disclosing the attribute value itself. However, current Anonymous Credential Systems implementations such as Idemix are complex and difficult to manage by final users.

Identity Management based on *Self-Sovereign Identities (SSI)* systems [56] [57] focus on providing a privacy- respectful solution, enabling users with full control and management of their personal identity data without needing a third-party authority taking over the identity management operations. Thus, citizens are not anymore data subjects; instead, they become the data controller of their own identity as they can determine the purposes and ways in which personal data is processed, as they manage directly their personal data during their online transactions.

## 2.6 Powerful user authorization and access control

In all security initiatives, ensuring who gets access to what in a legal, controllable, proportional and auditable manner is key. Limiting access to resources by establishing permission rules provides for better control over users' actions. Authorisation should be granted on the principle of least privilege, granting no more privilege than is required to perform a task/job, and the privilege should not extend beyond the minimum time required to complete the task. This restrictive process limits access, creates a separation of duties and increases accountability. *Privileged Access Management (PAM)* should be a tool used to control who accesses data and the systems being accessed.

While the cookie-based authorisation is used as the de-facto standard for communication between client server applications, this technology cannot be used when we have a multi- domain networks, taking this into account a new token-based technology (RFC7519) [58] was proposed that allows to make AJAX/REST calls to other domains, by including the user information in HTTP header of the http request. This authorisation mechanism cannot cope with the heterogeneity and diversity of health and care specific IoT devices and applications.

*Enrolment over Secure Transport (EST)* [59] is a protocol for bootstrapping certificate and the associated *Certification Authority (CA)* certificates over TLS and HTTP. The *IETF Autonomic Networking Integrated Model and Approach (ANIMA)* working group uses EST for a solution for automated *Bootstrapping Remote Secure Key Infrastructures (BRSKI)* [60], using certificates that are conceived for large scale, but it is not considered for constrained devices.

The OAuth is a delegated authorization framework enabling secure authorization for applications on top of the transport layer i.e., http-over-TLS. It allows the client-server applications to communicate and exchange data in a way to prevent eavesdropping and tampering [61] . Several works have been done regarding the OAuth security and its application in constrained environments. The scheme [62] focuses on using OAuth to allow only authenticated users to access the IOT network. Tysowski [63] discusses use of OAuth 2.0 for secure authorization of services running on different platforms. It discusses how OAuth can be combined with Open ID to provide both authentication and authorization. Solapurkar [64] proposes a scheme based on OAuth 2.0 and JSON Web Token [58] for securing existing health care services in the IOT cloud platform.

Alternatively, access control and authentication in BAN enabled devices has been proposed to be done using distance bounding protocols [65] [66] including anomaly detections in the wireless channel [67]. However, distance bounding by itself provides for only weak authentication and can be compromised by various attack techniques [68] (like the replay attack of Kfir et al. [69]). Further, authentication can be also done using out-of-band channels like audio or visual channel signals outside the standard communication channel [70].

## 2.7 Efficient and secure cryptographic mechanisms for data sharing and analysis

Healthcare organizations collect both administrative and clinical data relevant to support and improve the wellness, health and healthcare of individuals. The importance of these data depends on the users who have authored them. Article 32 of the GDPR states the encryption of personal data as a way of ensuring data security.

End-to-end encryption of shared data, in transit and in rest, while maintaining usability and efficiency on the end-user side is an open research challenge that still needs to be covered effectively to protect user's privacy. In this sense, new techniques, algorithms and protocols, e.g., those based on proxy re-encryption, are needed to reinforce security/privacy while outsourcing the computation to Cloud wallets to minimize user's risks in protecting crypto- material. In addition, new crypto-privacy techniques are needed to guarantee authenticity on the data through novel signatures schemes. Securing a medical device from software and hardware attacks is a demanding problem due to its small processing power and low power consumption requirement (BAN based devices) or its proprietary software.

One of the most popular cryptographic mechanisms involves the use of *Attribute-Based Encryption (ABE),* which utilises a public key cryptography to eliminate unauthorised data access in the cloud [71]. ABE encryption is employed in a number of eHealth system architectures [72] [73] [74] [75] [76]. Other mechanisms are the *Identity-Based Encryption (IBE)* [77], which places emphasis on the encryption of the data source and *Homomorphic Encryption (HE)* [78], which allows calculations to be performed on encrypted data without decrypting it first.

However, cryptographic schemes (especially public key cryptography) used in security apps are computationally and memory demanding. Lightweight cryptography schemes can be used to solve this [79] but however do not offer very strong security. Non lightweight solutions, such as EST over secure CoAP [80] propose an adaptation of EST for constrained devices, i.e., IoT devices, that work on top of *Datagram Transport Layer Security (DTLS)*. This has the consequent limitations that some devices will lack the resources to handle large payloads managed in EST-coaps.

On another note, all exchanged data should be encrypted, without intermediate entities such as proxies or cloud-providers being able to access the user's data. Data minimization and privacy- by-default properties, above all, in emerging distributed deployments needs to be guaranteed. Thus, novel cryptoprivacy protocols, mechanism and systems, such as those based on Zero- knowledge proofs, are needed to ensure anonymity, minimal disclosure of personal information, above all in public Clouds, ledgers and mobiles, while ensuring the user's rights laid out in GDPR.

Health data analytics, one of the EMD provided services, raises new concerns about privacy preservation, as the possible dynamic combination of large data coming from diverse sources can undermine anonymity, pseudonymity properties that can be given for granted in a single domain. The MapReduce framework is one of the best approaches used by industry leaders to implement cryptographic security in large datasets [81]. Currently, HE is also being exploited for health data analytics to perform computations on encrypted data without compromising patient privacy [82].

## 2.8 Deter unethical use of health information via audit trails.

Audit trails, or records of information access events, can provide one of the strongest deterrents to abuse. Audit trails record details about information access, including the identity of the requester, the date and time of the request, the source and destination of the request, a descriptor of the information retrieved, and perhaps a reason for the access. The effectiveness of such a record depends on strong authentication of users having access to the system. Audit trail information must also be kept in a safe place so that intruders cannot modify the trail to erase evidence of their access. Finally, although there is some benefit in users' *thinking* that an audit trail is being kept and analyzed, such trails are truly effective only if their information is *actually* reviewed and analyzed.

Effective software tools are needed to maintain continuous surveillance of audit trail information so that abuses are detected quickly and sanctions meted out, both to maintain the effectiveness of audit trails as prevention tools and to contain, as soon as possible, the extent of any abuse. Blockchain [83] is the best-known distributed ledger technology; a ledger is a database which keeps a final and definitive record of transactions. Records, once stored, are immutable and cannot be tampered without leaving behind a clear track. Blockchain enables a ledger to be held in a network across a series of nodes, which avoids one centralised location and the need for intermediaries' services. This is particularly helpful for providing trust, traceability and security in systems that exchange data or assets. Blockchains have found applicability in sectors ranging from banking and finance to public services and healthcare [84].

Gipp et al. [85] presented a decentralized trusted timestamping system. It enables users to prove that they were in possession of a file at a specific point in time in the past. Users have to hash the file and embed the produced hash value in a bitcoin transaction. The integrity of the data is ensured by the blockchain; checking the public ledger of transactions to check the validity of the proof of possession is trivial. Dot-bit [86] is a decentralized domain-name registration service that practically runs on Namecoin [87], one of the first forks of bitcoin. Dot-bit replaces domain-name controllers by a public ledger that moves domain-name registration records from the servers to the clients. Controllers as single points of failure are eliminated, while many web attacks targeting DNS servers become irrelevant. The authors in [88] implement a decentralized *Public Key Infrastructure (PKI)* service. Their system, called Certcoin, has no central authority, and requires the use of secure distributed dictionary data structures in order to store data related to the keys. Certcoin is implemented on bitcoin infrastructure. Recently, the connection of the IoT to blockchain has drown significant attention, mainly because of a relevant joint research between IBM and Samsung [89]. ADEPT is a system that uses elements of bitcoin's underlying design to build a distributed network of devices, – a decentralized Internet of Things. It taps blockchains to provide the backbone of the system, utilizing a mix of proof-of-work and proof-of-stake in order to secure transactions. IBM and Samsung chose three protocols – BitTorrent (file sharing), Ethereum (smart contracts) and TeleHash (peer-to-peer messaging) – to underpin the ADEPT concept. In [90], a peer-to-peer secure communication system is described for the IoT. The authors try to limit the required computational and communication effort for the nodes, while keeping the security requirements satisfied. However, even though they protect the communication, the storage of information remains unprotected. The protocol presented mainly decentralizes storage, in order to diminish the negative effects of centralized storage schemes. Another relevant approach is presented in [91], wherein a heterogeneous network infrastructure is designed, to allow different applications in the IoT to interact with sensors and actuators.

## 2.9 Encryption of communications

Each type of external access to health care information resources poses possible security vulnerabilities that could compromise patient privacy. One area where recent advances has been made to improve the security (and arguably also privacy) of users is in the encryption of communications. Some recent advances have made such *End-to-End Encrypted (E2EE)* messaging a commonplace and easy to use experience for hundreds of millions or even billions of users.

The important technology behind this development is the Signal protocol by Open Whisper Systems [92] and the Signal application [93]. When WhatsApp adopted this technology to provide E2EE messaging for their users, it signaled a major change in the encryption landscape of communications between individuals. Of course, there are many other applications that now provide similar protection of communications and even Facebook is said to be contemplating adding E2EE messaging to their Messenger application e.g., [94]. Several examples [95] show, that it is possible to achieve great security benefits through secure messaging, without affecting user experience in any meaningful way.

E2EE is not the only field where additional security can be achieved through encryption. Network traffic has been largely unencrypted until some recent developments that have produced website developers' easy tools to make their sites run HTTPS, the encrypted version of HTTP. The Let's Encrypt -project provides an easy way to secure your website and their statistics show a remarkable increase in HTTPS adoption [96]. HTTPS has been available for a long time, but the setting up of a certificate and all other setup for the encryption has been hard for the administrators.

For the end users, many browsers offering functionality that will enforce HTTPS is used in browsing whenever possible [97]. This makes the user experience very smooth also for the web end users. Of course, this type of encryption brings some side effects such as that users tend to ignore security warnings (e.g., because a certificate has expired) and thus can be exposed to phishing etc. [98]

To further protect wireless medical devices, external modules mediating communication with the Medical device and providing both confidentiality for transmitted data and protection against unauthenticated communication, have been proposed. Such devices are the cloaker [99], the IMDGuard [100] or use friendly jamming tokens as the ones in [101].

## 2.10 Compliance to legal and regulatory frameworks and standards

Legal and regulatory developments related to cybersecurity are increasing in the last decades. At European level, GDPR (General Data Protection Regulation) [12] and ISO 27001 [102] are two important compliance standards that aim to strengthen data security and reduce the risk of data breaches. GDPR is a regulation in EU law that regulates how companies process and protect personal data relating to individual citizens in the EU. ISO 27001 is an international management standard that provides a proven framework for managing information security. It uses an integrated set of recommended policies, procedures, documents and technologies in the form of an Information Security Management System.

Patient safety risks related with medical devices or software, are typically managed through specific frameworks which focus on the device/software to be developed. For example, ISO 14971 [103] provides a Risk Management Framework designed for the development of medical devices. Furthermore, ISO/TR 27809:2007 [104] provides guidance regarding patient safety in the context of the ISO 27000 family of standards related with software security. More specifically, ISO TR 27809 identifies specific controls which can be used as guide to identify and manage possible Patient Safety risks related with the Medical Software security issues. Moreover, European Commission provides a regulation framework regarding Medical Device development consisting of several directives regarding application of security measures, the application of CE marking etc. [105]. Respectively, FDA provides guidelines on post market management of Cybersecurity in Medical Devices [106], also focusing on preventing patient harm. The above regulation and technical risk management frameworks and standards, are characterised by the following features:

a) They focus on the industry perspective and tend to provide guidance on how to manage risks from a manufacturer point of view and do not actively engage healthcare professionals who are the main people responsible for the overall Patient's Safety;

b) They do not emphasize on the healthcare setting context which is volatile and subjective (e.g., prioritisation of needs and risk-benefit relationships is heavily dependent on specific patient and healthcare process);

c) They define rigorous and hard to adopt risk management approaches which typically refer to cybersecurity experts and therefore are not easy to apply in a flexible manner in a real- world healthcare context.

## 2.11 Related EU projects

In the EU-funded research community there are many related research projects in terms of the EMD security and trust. ASCLEPIOS [107] seeks to maximize and fortify the trust of users on cloud-based healthcare services by developing mechanisms for protecting both corporate and personal sensitive data. While researchers have developed many theoretical models that could enhance the security level of healthcare services, only a rudimentary set of techniques are currently in use. ASCLEPIOS is exploiting this gap by using several modern cryptographic approaches to build a cloud-based eHealth framework that protects users' privacy and prevents both internal and external attacks.

MyHealthMyData (MHMD) [108] aims at fundamentally changing the way sensitive data are shared. MHMD is poised to be the first open biomedical information network, centered on the connection between organisations and individuals, encouraging hospitals to start making anonymised data available for open research, while prompting citizens to become the ultimate owners and controllers of their health data.

MHMD is intended to become a true information marketplace, based on new mechanisms of trust and direct, value-based relationships between EU citizens, hospitals, research centers and businesses.

PANACEA [109] is driving a people- centric approach to cyber security in healthcare. Running from January 2019 to December 2021, this research and innovation action will design, develop and deploy the PANACEA Toolkit for uptake in hospitals, care centres and other medical facilities.

FeatureCloud [110] is a transformative, pan-European research collaboration and AI-development project which implements a software toolkit for substantially reducing cyber risks to healthcare infrastructure by employing the worldwide first "privacy by design" approach.

KONFIDO [111] is a H2020 project that aims to leverage proven tools and procedures, as well as novel approaches and cutting-edge technology, in view of creating a scalable and holistic paradigm for secure inner- and cross-border exchange, storage and overall handling of healthcare data in a legal and ethical way both at national and European levels.

The Serums Project [112] deals with security and privacy of future-generation healthcare systems, putting patients at the centre of future healthcare provision, enhancing their personal care and maximising the quality of treatment they receive.

SPHINX [113] aims to introduce a Universal Cyber Security Toolkit, thus enhancing the cyber protection of Health IT Ecosystem and ensuring the patient data privacy and integrity. It will also provide an automated zero-touch device and service verification toolkit that will be easily adapted or embedded on existing, medical, clinical or health available infrastructures.

CUREX [114] goal is to safeguard patient privacy and increase their trust in the currently vulnerable critical healthcare information infrastructures, especially in cases where data is exchanged among healthcare stakeholders within any business, operational and systemic cross-border environment.

SHiELD's [115] vision is to unlock the value of health data to European citizens and businesses by overcoming security and regulatory challenges that today prevent this data being exchanged with those who need it. This will make it possible to provide better health care to mobile citizens across European borders, and facilitate legitimate commercial uses of health data.

STORK 2.0 contributed to the realization of a single European electronic identification (e-ID) and authentication solution. It built on the results of STORK, establishing interoperability of different approaches at national and EU level, eID for persons, eID for legal entities. The results of the STORK project are currently available via the CEF Digital Portal [116].

Finally e-SENS [117] focuses on cross-border interoperability in eHealth, eJustice and eProcurement, aiming to provide generic and re-usable software components for, inter alia, e-Delivery, e-Identity (eID) and e-Signature. Among other tools they provide a standalone adapter in the e-ID area to bridge the gap between e-IDAS based German middelware and the Dutch PEPS based on STORK 2.0.

## 2.12 Innovation potential for cybersecurity in the healthcare sector.
### 2.12.1 Healthcare data breaches

Data breach refers to the intentional or unintentional release of secure or private/confidential information to an untrusted environment [118]. More specifically, a medical data breach is a data breach of health information, which could include either the personal health information of any individual's electronic health record or medical billing information. Among 3,824 data breach notifications reported between May 2018 and February 2020, 244 (6.4%) is related to the health sector, increasing by a factor of four between 2018 and 2019. Data breach characteristics of the health sector were similar to data breach characteristics of the other sectors. Loss of confidentiality is the most important breach (80.7%) in the health sector, followed by the loss of availability (27.5%); some data breaches are mixed. 175 (71.7%) notifications reported fewer than 300 people impacted. The malicious cause occurred in 58.2% of them, and accidental cause accounted for 25% [119]. These security violations are generally the outcome of other threats endangering the health organizations' ICT infrastructure, including Electronic Medical Devices. Concerning this great challenge, ***HEIR offers a solution to reduce the detection time of data breaches and increase security monitoring accuracy, which leads to cybersecurity incident investigations resolved within an acceptable timeframe for the organizations.***

Providing advanced security services is the subject of many innovative initiatives nowadays. Machine learning is changing organizations' approach to threat detection and how they adapt and adopt cybersecurity processes. The idea is not just to identify and prevent threats but to mitigate them as well. An algorithm can learn from its mistakes on the fly. It is always the best version of itself because it is continuously improving its performance. A good ML discipline is one that can "see" patterns of behavior, guessing the form of an attack and how to fight back. The algorithm can be trained with different types of attacks, learn the methods to gain privileged access and lateral movements, and even adapt in real-time to a situation. An excellent ML approach can learn from false positives. False positives will always exist, but they are reduced with each interaction with an algorithm because the machine is continuously learning. After implementing an ML system, false positives can be reduced by 50% to 90% [120]. Google is an indicative example as it is officially expanding its Chronicle cybersecurity platform into the threat detection realm: by placing ML algorithms into Chronicle, which can then analyze vast swaths of data, this enables the system to identify security threats more quickly – Google has set the wheels in motion for proactive threat detection and alert functionality. At the baseline of the solution, there is also an intelligent data fusion, combining a new data model with the ability to automatically connect multiple "events" into a single unified timeline [121].

As far as secure data sharing between healthcare entities are concerned, scientists envision safe, collaborative infrastructures. As reported in relevant research [122], such an infrastructure could be based on blockchain. However, while blockchain is a possible solution to secure the health data of patients, the question is whether the technology is too early in its infancy or if the cost to set up the infrastructure is too high at this moment in time. Of course, the most critical hurdle of all is implementing this technology within the parameters set forth by regulators in the healthcare space. These are the challenges which ***HEIR focused to tackle through a threat identification and cybersecurity knowledge base system that supports trustworthy data exchange across the healthcare supply chain, threat prevention, detection, mitigation, benchmarking, and certified assurance.***

2.12.2 Healthcare ICT: Infrastructures, systems, services, medical devices etc.

Recent events have shown the impact of cyber-attacks on infrastructures that we do not expect to fail. The Wannacry incident [123] has demonstrated the increasing reliance on medical devices on classic network and *Information and Communication Technologies (ICT)*. Such technologies, while bringing increased facilities and opportunities, also opens the door to vulnerabilities and attacks. Traditional operation of these medical devices assumed that they would not be accessible from the outside world and would not be connected to the hospital's network. The need to transfer information from the devices to medical files to facilitate care, and the need for these devices to be accessible from the outside world for maintenance, have opened an entirely new set of vulnerabilities. Several research and innovation topics are relevant for this challenge.

***Identification and authentication, specifically dedicated to the needs of healthcare personnel.*** The specific aspects of healthcare operations, such as 24x7 care for patients, increasing use of remote care where you can leave the monitoring devices in the hands of patients (telehealth medicine), emergency situations requiring immediate access without barriers, are usually not well accommodated by classic *Identity and Access Management (IAM)* solutions available today. Healthcare aspects of security policies have been studied in research literature, but there are no practical implementations available today. Furthermore, organizations tend to rely on well-known tools (e.g., office tools, disk and printer sharing, etc.), where end-users are familiar with the technologies and operators skilled in operating these systems are available. This leads to an additional acceptance and training challenge for the personnel who has to use these new tools.

***Definition of monitoring and detection strategies for healthcare networks, complementing access control to highlight anomalous network behavior.*** Anomaly detection is hard to establish within healthcare environments, as operating conditions may vary due to the type and complexity of the host institute. For example, EMDs can be placed in patient homes, thus lying outside the classic security monitoring domain. Private "medical" devices are also brought on premises, as well as other equipment such as smartphones and connected watches.

***Definition of mitigation strategies.*** Based on the detected anomalies, the security system in place should deploy flexible network overlays to verify and enforce network activity compliance with security policies. Mitigation remains a difficult topic, particularly with automation, as network management operators are reluctant to let automated processes take the upper hand when dealing with cyber-attacks.

***Deployment of these technologies in the context of the cloudification of network infrastructures.*** Healthcare environments, like others, are moving towards the cloud. This induces a transfer of control from healthcare organizations to external service providers. On one hand, this may help these healthcare organization get better cybersecurity support. On the other hand, this may create additional difficulties ensuring the protection of sensitive processes and information. Understanding how cloudification of healthcare infrastructures impacts cyber- risk management is a forthcoming important research and innovation challenge.

***Internet of Medical Things (IoMT)***. The medical world is increasingly relying on devices to sense, measure, and produce data. These devices are extremely heterogeneous. Some of them, like imaging devices, are large and computationally powerful, but also very expensive and sensitive to ICT-related problems such as loss of connectivity. Other, smaller devices, like heartbeat monitors or blood glucose measurement devices, are remote and can be handled by untrained users. All connected devices forming the *Internet of Medical Things (IoMT)* are extremely sensitive to cyber-attacks. The extreme heterogeneity of end user capabilities is thus an additional challenge, requiring both adaptation of existing technologies, development of new intuitive HMI systems to manage them, increase in the capability of these IoMT devices to defend themselves autonomously, and innovative user training and feedback.

### 2.12.3 Privacy-sensitive data

New types of pervasive wearable technology bring interesting insights into personal daily life and facilitate various health-related activities. The wearables market growth is stimulated by miniaturization and the development of new types of sensors used by device vendors to create new consumer-level kinds of devices. The gathered data is often transferred to a smartphone- centered ecosystem that provides user-friendly interfaces for visualization, notifications, and further data sharing. Examples of such devices are fitness trackers, smartwatches, smart rings, or smart glasses. Complementary privacy policies are, however, often provided in a hard-to- understand way without clear answers to fundamental questions such as: *"What data is being collected?"*, *"How is the data protected?"* and *"Who has access to the data?"*. Also, since the device's advertised functionality could be based on secondary processing of more primitive data collected from various sensors, it might be difficult for a consumer to relate to all these three privacy dimensions.

By not accepting the provider's data security and privacy agreement, a user is often simply not able to use a full portfolio of the device's features or sometimes not able to use the device or service at all. On the other hand, by accepting it, a user gives consent for various data-transfer activities happening in the background, and often even the ownership of her/his data. Numerous security analyses have been performed to investigate whether wearable devices, smartphone applications, and associated cloud services fulfil the current rules for data security and privacy requirements. Even though these efforts often conclude with alarming findings, this has not yet led to much change in vendors' practice in general when it comes to data security and privacy- related ethical issues, as amassing large volumes of human health data can enable the development of more refined and predictive software which could be further monetized. Whilst the motive for additional post-processing might not be apparent to the device's end-user at first sight, the uncovered insights provide a potentially compelling advantage over the rest of business competitors in various areas such as targeting advertisement. Therefore, secondary data usage scenarios include data reselling to third-party marketers and insurance companies. Another purpose of this data analysis is usually marketed as product improvement. However, since metadata collection uses additional processing or network resources, it may negatively influence user's experience with the product in various aspects.

As a consequence of this vendor-based user data harvesting, new regulations such as the General Data Protection Regulation [12], have been introduced to *provide increased legal certainty for both individuals and organizations* [124]. After introducing GDPR, it is now common to see applications integrating an explicit user- adjustable set of options dedicated to data collection related settings. This is usually hard to understand and adjust by the common user, resulting in it being generally ignored by the user choosing "select all" or similar – not knowing what (s)he is accepting all these blanket consent options. Recently, multiple companies have provided options to give end-user consent (opt-in) or withdraw (opt-out) for the purpose of further analysis. Interestingly enough, the existence of regulations such as GDPR enforces regulatory compliance and, indirectly, influences the technology stack behind the technical solution. For example, GDPR-incorporated right to erasure, which is also known as 'the right to be forgotten', essentially imposing design update of distributed Blockchain platform to address all of the GDPR concerns fully. Based on the described development in this area, multiple challenges need to be continuously addressed:

- coherent way of presenting and interpreting privacy policies to the end-user,

- developer's guidance on implementing opt-in/-out option in various applications, in a way that is easy to relate to for the end-users,

- up-to-date framework that describes technical prerequisites in relation to current regulatory requirements,

- user-friendly ways for users to access, manage, and delete their health data.

### 2.12.4 Legacy systems

For this aspect, we will provide as an example the situation in the UK. In 2015 the UK NHS committed to migrating from paper records to *Electronic Patient Records (EPRs)* by 2020. All hospitals should comply with the aforementioned direction. For instance, Croydon, as an integrated health care system, with links to Primary Care, via Co-ordinate My Care means that Croydon NHS Trust needs to be able to integrate primary care records to EPR. This is a major problem given that the Primary care systems operate using different medical applications. Beyond this complexity of merging differing application, then further research is needed to ensure seamless updates and synchronization of disparate systems so as to avoid duplication and erroneous correction of entries due to differing time stamps. A hospital-based EPR, enables the Primary care to see records through a portal. However, for the patient who is mobile, and moves out of their local catchment area, then their health records, currently, do not follow them easily. So, information may be lost to third parties beyond the reach of the hospital-based EPR. Truly mobile methods to enable the patient to see their records and to access their data are of high priority. The development of the NHS application by NHS digital is a means to address that.

Third-party organizations, such as *The Patient Knows Best* [125] and Clevermed [126], have developed a Web-based system whereby patient details are located in central servers, but access is via a web-based application and can enable notes to be accessed anywhere. The NHS currently purchase these systems on an ad-hoc basis: it is not uniformly rolled out. Maternity and Oncology services currently have patient-held paper records, but some records are also migrating to web-based – the maternity system is one such migration. Again, this strategy is part of the NHS ethos, but currently not there yet: this is due to the NHS's reliance on third parties to develop the software for the NHS to use.

### 2.12.5 User awareness

A significant barrier for adopting and exploiting advanced technologies to address security and privacy challenges in healthcare environments is the lack of user awareness. One of the less discussed aspect of this problem is how standards can play an important role in improving approaches to information/data security across different geographical regions and communities and also promote the successful acceptance of best practices in cybersecurity /privacy and personal data protection. HEIR worked together with relevant standardization bodies and CERTS all over the EU, exploiting FORTH's collaborations. HEIR's goal was to extend state of the art to a set of new standards that can upgrade medical IT applications' security to a satisfactory level. The identified set of standardization bodies and EU directives that were identified, contacted to the extent possible and closely monitored during the project lifetime, include:

**The NIS Directive:** the EU directive aims to create and strengthen a *Computer Security Incident Response Team (CSIRT)* to promote cooperation between all Member States (MS) and create a culture of security across sectors such as digital infrastructure, manufacturing, transport, energy, healthcare, financial market, water.

**The European Union Agency for Cybersecurity (ENISA):** ENISA is a center of cybersecurity expertise in Europe and supports MS for more than ten years in implementing relevant EU legislation.

**The EU Cyber Security Strategy:** this strategy provides a harmonized framework for the evolution of three different cybersecurity aspects, which until recently had been evolving independently.

**The EU Cloud Strategy:** the European Commission (EC) published its cloud strategy, entitled 'Unleashing the Potential of Cloud Computing in Europe.'

**ETSI Cyber Security Technical Committee (TC CYBER):** is recognized in the EU & worldwide as trusted experts offering market-driven cybersecurity standardization solutions and guidance. TC CYBER is working closely with relevant stakeholders to develop appropriate standards to increase privacy and security for organizations and citizens across Europe.

**CEN-CENELEC-ETSI 'Cyber Security Coordination Group' (CSCG):** The group aims to provide strategic advice in the field of IT security, *Network and Information Security (NIS)*, and cybersecurity.

**HIMSS Europe** that is a leading health IT knowledge organization that acts as a barometer for the industry and provides valuable insights into market trends and gap analysis at a local, national and international level.

**NIST Cybersecurity Framework and ICS-CERT medical cybersecurity advisories**: HEIR will collect and share information about medical objects' vulnerabilities and the appropriate mitigation actions.

Based on the aforementioned standardization bodies, the project targeted specific standards to evaluate and examine possible contributions:

**Information Security Standards:** *ISO/IEC 27001:2013* Information security management systems (ISMS), *ISO/IEC 27003:2010* Information security management system implementation guidance; *ISO/IEC 27005* Information technology — Security techniques — Information security risk management; *ISO/IEC 27014:2013* governance of information security; *ISO/IEC TR 27016:2014* Information security management - Organizational economics, *ISO/IEC 27039:2015* Selection, deployment and operations of intrusion detection systems (IDPS), *ISO/IEC 27040:2015* Storage security, *ETSI TR-103-305* Critical Security Controls for Effective Cyber Defense;

**Data Protection and Privacy Standards**: *ISO/IEC 27018:2014* Code of practice to protect personally identifiable information (PII) in public clouds acting as PII processors, *ISO/IEC 29100:2011* Privacy framework, *ISO/IEC 29101:2013* Privacy architecture framework, *BSI BS 10012:2009* Data protection. Specification for a personal information management system, *CEN CWA 16113:2010* Personal Data Protection Good Practices

**Third Party Security Management Standards:** *ISO/IEC 27036-1, 2 and 3:2014* Information security for supplier relationships - Parts 1, 2 and 3, *ISO 28000:2007* Specification for security management systems for the supply chain

Apart from the standardization activities, HEIR focused on raising cybersecurity awareness to executives and employees in the healthcare sector, defining the duties, responsibilities, and communication procedures and protocols of all the members, ensuring at the same time alignment with current directives and legislations; thus, significantly advancing Security Governance in the health sector. Moreover, the HEIR Observatory for the Security of Electronic Medical Devices (OSEMD) became a cybersecurity and resilience benchmarking tool for medical IT devices, networks, and computer services, acting as a public repository for best practices and solutions towards healthcare cybersecurity as well as a monitoring service for cybersecurity issues in the medical sector. This enables healthcare stakeholders to safely access, monitor, and share information about HEIR good practices/successful scenarios and mitigate identified challenges, problems, and vulnerabilities. More than that, analytical guidelines and recommendations were prepared and presented to enable the EU to start a fruitful discussion on the necessary directives that should be introduced to be applicable to software being developed for medical systems. In multiple cases, the security of electronic devices and applications in the healthcare ecosystem is more of an organizational and procedural challenge rather than a technical one. To complement all the above activities, HEIR project built a large community and an ecosystem around the project's results and impact assessment outcomes that can promote public awareness for security and privacy in the healthcare domain. This community was nurtured through networking and liaisons with technical and domain-specific communities, policymakers and local authorities, EU associations (i.e., EuroVR, BDVA), other EU projects, and more.

### 2.12.6 Trust increase

The Health Care Industry Cybersecurity Task Force report [127] also provides two imperatives to Develop the health care workforce capacity necessary to *prioritize and ensure cybersecurity awareness and technical capabilities* (*Imperative 3*) and to *increase health care industry readiness through improved cybersecurity awareness and education* (*Imperative 4*). It is universally recognized that undertrained employees are organizations' biggest cybersecurity weakness. The problem is even more evident in healthcare since the operator's attention is attracted to her/his main priority, the patient's health. According to a study issued by the Ponemon Institute in 2016 [128], 36% of healthcare organizations that have been breached point to unintentional actions by their employees as the cause. In November 2016 alone, 54% of breaches were caused by employee error, a record month for breaches. So, similarly to proper hospital hygiene practices, cybersecurity cannot become a common practice without training and education. For example, another survey [129], addressed to qualified employees of providers and payers in the United States and Canada, reports that 21% of healthcare employees write down username and password near the computer. An effective training model should teach them to avoid this behaviour.

There is broad evidence [130] [131] that security awareness training is the most cost-effective form of security control. On the other hand, a meaningful approach to the training cannot be based only on the transmission of technical and legislative information and you have to consider which is the perception of the risk as seen from a psychological point of view [132] [133]. The problem has been around for some time and, currently, there are a lot of training courses available in the market with most of them being supplied online. Some of them stimulate interactivity, a fundamental element for the training's success [134]. However, they are often focused on specific threats, like phishing, with marginal care on other aspects, which are equally important, especially in the healthcare environment. Because of the subject's practical relevance, organizations such as SANS are involved in the definition of training programs and certifications, taking into account the specific area of healthcare professionals. However, the major part of the training programs takes as a reference point, the US scenario, where the legislative and organizational framework is often different from other regions like e.g. Europe.

Key points in the training are the awareness and human behaviour of people involved, mainly focusing on cybersecurity assurance. In the healthcare sector, continuing education and professional development are crucial to maintain and advance skills and knowledge in an environment that is continuously changing due to new healthcare research and technology. E- learning provides a solution to these challenges, allowing healthcare professionals to follow the training at their own pace at a time and location that suits them. Sometimes online training does not suffice, though (e.g., when lab training is necessary), in which case a hybrid solution (known as blended learning), where part of the course is delivered through classroom lectures, can exploit the best of both worlds. The ability to include interactive and multimedia elements in e- learning is also vital because it can help improve retention and understanding of medical course material that is often very visual. Indeed, the result of a study conducted on these issues [135] indicated that different delivery models should be used together to get the maximum benefits out of the information security awareness program.

Finally, another appealing opportunity is found in using gamification. By this term, we mean "the use of game design elements in non-game contexts". Gamification is often employed in health and wellness apps related to self-management, disease prevention, medication adherence, medical education-related simulations, and some telehealth programs [136] [137] [138].

# 3  Architecture

## 3.1  Design Methodology

As already stated in the introduction, HEIR was developed to offer a holistic healthcare-oriented, cybersecurity platform, incorporating threat hunting, ml-based anomaly detection, and privacy-aware capabilities. The architecture of our solution is presented in Figure 3.

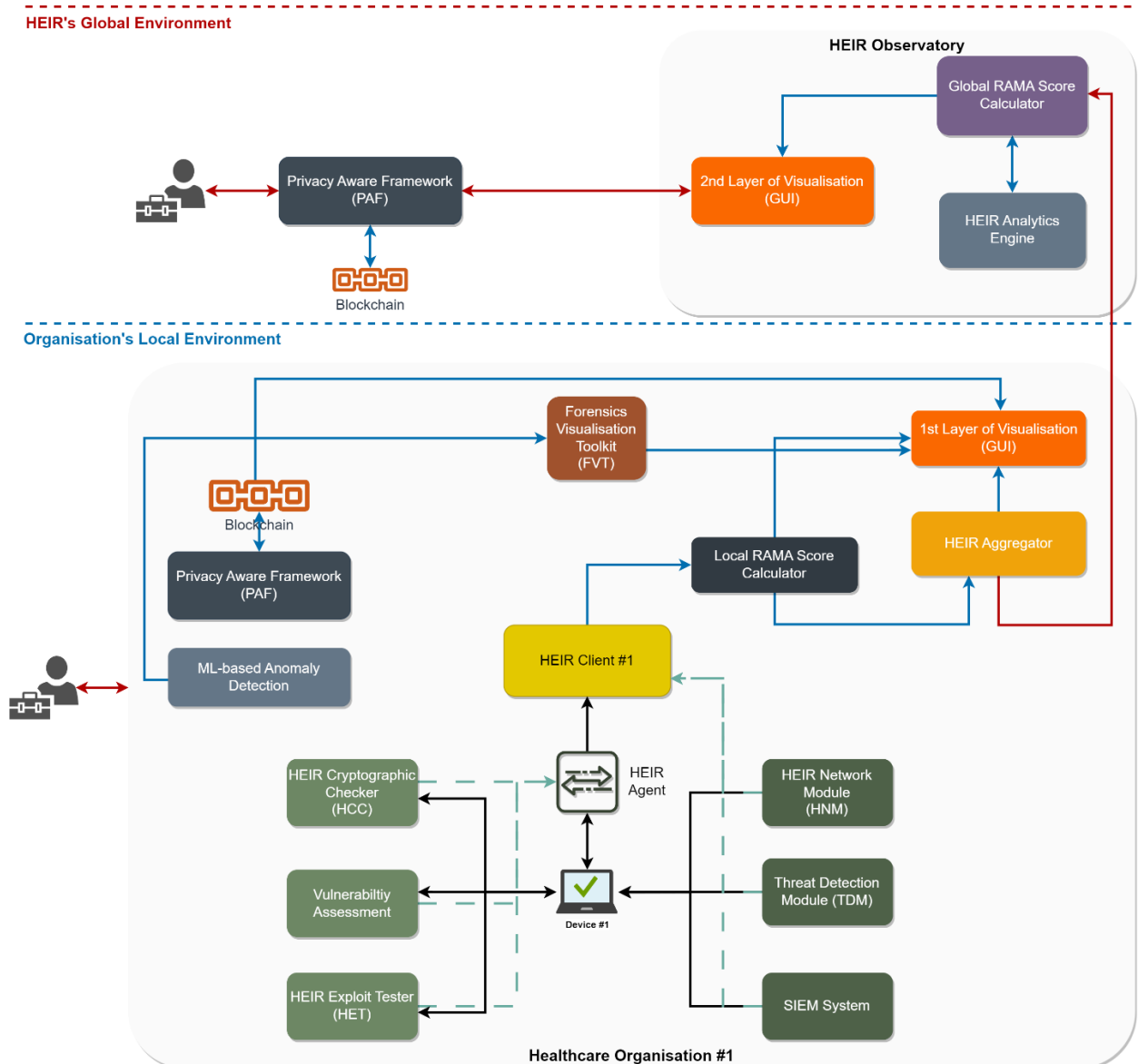

*Figure 3: Architecture Diagram*

## 3.2 A snapshot of the offered solution

HEIR framework is based on a multi-layered hierarchical architecture. It comprises of several risk assessment tools that communicate with the HEIR agents and clients, operating inside a healthcare facility, and provide the output to the Local RAMA score calculator. Apart from the local analysis, these clients submit anonymized findings to the Global RAMA score Calculator, through the HEIR aggregator, where they are being visualized in HEIR's Observatory.

The HEIR framework follows a modular architecture which can be modified to support different and more complex healthcare environments. It can also be further extended to support new threats and provide additional recommendations. One critical component for this hierarchical architecture is the HEIR Aggregator. In large healthcare environments as a hospital with many departments, different medical devices, and subnetworks, a single Local RAMA score may not be enough to support the IT administrators in understanding all the necessary details for every department. The HEIR Aggregator is there to collect the data from all the Local RAMA score calculators, aggregate them and finally provide detailed feedback to HEIR's 1st Layer GUI. The aggregated information is also anonymized and further transmitted to the Global RAMA Score Calculator to extract the cybersecurity and resilience benchmark score of the whole organization compared to the global trends documented from other organizations.

### 3.3 Architecture Overview

The presentation of the HEIR solution architecture follows a logical path, from the use cases supported by the solution to the functional conceptualization of the needed components, and ultimately to the software modules and tools that can be used for the implementation of these functionalities. The architectural design aims to cover all these intricacies, while maintaining the openness of the system and ensuring that it will be scalable and easily modifiable. Furthermore, all the design and implementation decisions are grounded in established technology and industry standards.

We consider a design as successful when it covers all the following aspects:

- Usability
- Performance
- Security
- Maintainability
- Scalability
- Reliability

**Usability**: The architecture isolates the presentation components (1st and 2nd layer of visualizations), giving the possibility to focus on good User Experience design (UX). Thus, a web designer or a usability expert can work separately on the User Interface unaffected by the backend system developers. These experts can focus on the User Interface to maximize the quality of user experience. Internally we followed a Model View Controller design pattern for building the web application, starting from a plain, well-defined user interface that consumes the services provided by the backend. The use of modern web technologies for advanced visualisations (e.g. Hicharts.js [139], plotly [140] etc.) and interactive, responsive dashboards (e.g. React.js [141] or Angular 2+ [142]) offer the best set of Front-end features to provide a clean and fully functional interface. Finally, we followed an agile methodology for developing the HEIR solution, based on rapid prototyping and frequent iterations. This allowed for more frequent evaluations close to the end-user of the solution and better results in terms of meeting the usability requirements.

**Performance**: For tackling performance issues, we are relied on two factors – caching and distribution. The solution architecture logic is implemented in a modular way in the backend and communicates through Apache Kafka, an open-source distributed event streaming platform for high-performance data pipelines. Module services offer parallelization in calls to the backend and can be deployed independently in a distributed way, following a Software Oriented Architecture. If needed, load balancing and caching can be applied.

**Security**: The security features are applied system-wise, covering the whole architecture. An Access Control and Identity Management system ensures that only users with appropriate permissions will be able to access the data residing in the solution's data storage.

**Maintainability**: For addressing maintainability, we are following standards oriented and technology independent architecture. For communication between the various developer teams, we set in place Redmine [143] issue tracking system. This way all issues were traceable, and the history of the development process could be examined.

**Scalability**: The system can scale up based on the volume of the requests. The architecture is free to scale up easily due to the service oriented distributed nature of the backend. Scalability in terms of data is also desired for HEIR mainly because of potential big volumes of data that can be analysed and visualised.

**Reliability**: To build a reliable system, a certain number of characteristics were considered. These characteristics include maturity, availability, fault tolerance and recoverability as described in the software quality model of the BS ISO/IEC 25010:2011 standard. Some of the modules of the HEIR solution are based on existing solutions that have been used in the past and only required some adaptation to serve the needs of the users, therefore they are mature enough to be part of a reliable system. Furthermore, the modules that were created for the purposes of HEIR were also based on widely used technologies or pre-existing tools which could be easily supported by the owners or the community in the case of open-sourced solutions. The components of the solution were designed in a way that tries to help the end-users avoid mistakes and misuse of the system's functionality. Nevertheless, errors are always a possibility, so each component incorporates an internal error handling mechanism to be tolerant of misconfiguration or malicious input. Furthermore, the loosely coupled architecture of the solution avoids points of single failure and provides the ability to have a working production solution even if one of the modules temporarily fails to perform adequately. For example, temporary failure of a module would result into its output being unavailable but not in bringing the whole solution to a halt.

HEIR deploys agents (HEIR agents) in servers/workstations/devices to collect data from different layers. Wherever pertinent data can be obtained, such as Hospital Information System (HIS) data. Data deriving from the agents are portrayed on the 1$^{st}$ Layer GUI and the Forensics Visualization Toolkit (FVT) which serves as a threat hunting GUI for use by the hospital IT administrators. The risk assessment related data are fed to the HEIR Client processing system, and then to the Local RAMA score calculator to determine the RAMA score of an individual department. If the Pilot healthcare environment contains more departments, the individual RAMA scores per department are aggregated by the HEIR aggregator component so that a Hospital RAMA score is calculated. The above RAMA score and metadata is communicated to the Global RAMA score calculator, through HEIR's Aggregator, which is also communicated back to the hospital (1$^{st}$ layer GUI). Finally, the Global RAMA score, metadata, and recommendations are portrayed in HEIR's Observatory

## 3.4 Design and Deployment
The following deployment diagrams provide information about the framework deployment topology.

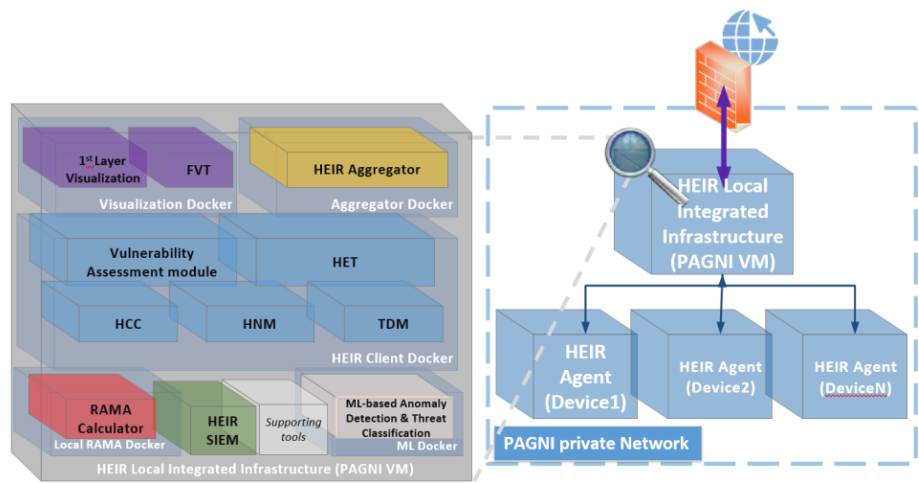

*Figure 4: Local deployment diagram*

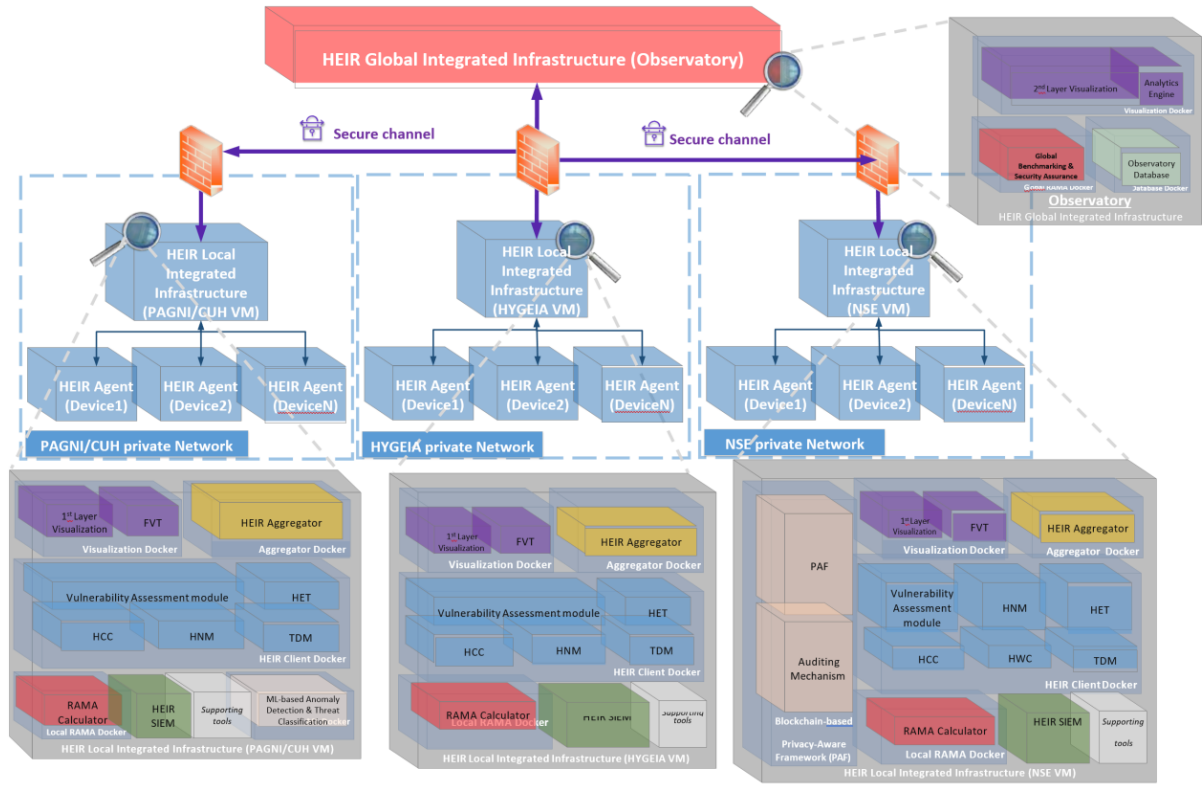

*Figure 5: Global integrated deployment diagram.*

The modularity of the HEIR integrated framework enables the use of tailored deployments of the framework per pilot. So, while all four pilots share the fundamental set of solution functionalities, different combinations of the HEIR individual modules were employed per pilot. So, in cases where relevant data is obtainable (such as the Pagni and the Croydon pilot sites), the ML component is utilized, while in NSE where there is the need to data is available the PAF component is employed. The deployment diagram of a local HEIR client (i.e., Pagni pilot site) is depicted in Figure 4. The global integrated deployment diagram of the HEIR eco-system containing multiple HEIR clients is shown in Figure 5.

### 3.4.1 HEIR Client

The HEIR Client collects and processes information, either on the endpoint level or centralized level. The architecture of the HEIR client is modular and can plug in several analysis components (e.g., HEIR Network Module, HEIR Cryptographic Checker, HEIR Exploit Tester etc). The HEIR Client architecture is depicted in Figure 6.

The tools send events with information about threats, security metrics, risks. This information is also received from the HEIR Agent of the facilitators package. The events are normalized and converted to a single and uniform format by the Event Taxonomy Translator component. The normalized events from different modules are aggregated by the Event Aggregator component which also correlates and processes them before forwarding them to other external components. In our example the external component is the Local RAMA score calculator. The events are submitted to the calculator by event publishing to Kafka Message Broker. Other modules can subscribe to the topics of interest.

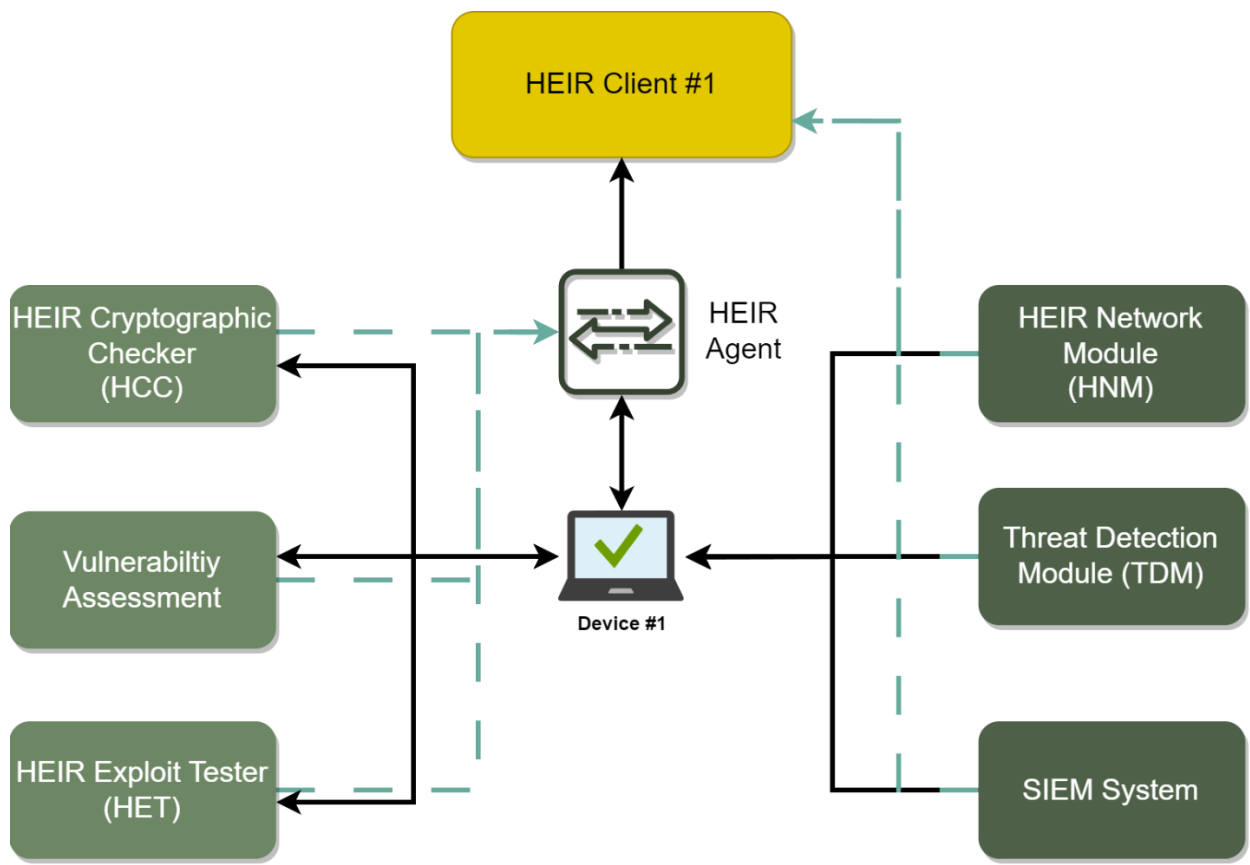

*Figure 6: HEIR Client Architecture*

### 3.4.2 Network module (HNM)

HEIR's Network Module monitors the assessed network traffic and provides security insights regarding any identified malicious activity. This module can analyze both the inbound and outbound network traffic of the system and detect (i) private information leaks, (ii) malicious content sent over the network, and (iii) ongoing attacks on the network. The output of this component provides connection information for the endpoints connected to the analyzed network and usage statistics that the anomaly detection component can use.

### 3.4.3 Exploit Tester (HET)

HEIR's Exploit Tester is responsible for assessing the attack surfaces for the operating system's configuration. Unlike the Vulnerability Assessment, which detects issues with the installed applications, the operating system's vulnerabilities and misconfigurations are discovered by the Exploit Tester. To accomplish that, it queries the OS's registry keys and configuration parameters as input, then outputs a list of the misconfigured security-related components, followed by suggestions and descriptions. Unlike the VA, the HET is focused on evaluating the attack surfaces based on operating system configurations. Some configurations can pose specific risks or might indicate some intrusion. For instance, the macro running enabled by default in Office applications increases significantly the attack surface, more specifically the risk of infection. Another example is disabling the task manager. It might be a legitimate action, but in most cases, it is disabled by malware or an attacker to make the process visibility harder and mitigate as well.

### 3.4.4 Cryptographic Checker (HCC)

The Cryptographic Checker (HCC) detects outdated security protocols within the assessed cyber system's servers or targets the external servers that are service providers for the system itself. HCC is based on SSLScan[1] and can detect the used protocol and its version and the usage of vulnerable cryptographic implementations. The former can be cross listed with the required ones per component (e.g., the latest TLS protocol).

### 3.4.5 Threat Detection Module (TDM)

The Threat Detection Module (TDM) is a module integrated technically in the HEIR Agent but provides information to the HEIR Client in the same way and with the same meaning as the HNM. The module is able to scan local files and/or processes from an endpoint machine and to detect malicious content when executed. It provides another layer of information about malicious activity that influences further the RAMA Score and alerts.

### 3.4.6 HEIR's Vulnerability Assessment

A Vulnerability Assessment (VA) is a systematic review of security weaknesses in an information system. It determines whether the system is vulnerable to known vulnerabilities, rates their seriousness, and makes remedy or mitigation recommendations. HEIR's VA module targets the Operating System (OS) configurations and application information. These points of interest might place the endpoint and the entire medical system at risk for security breaches if they are improperly set or if the applications are outdated. The module interacts with the assessed system and collects the application's name and version. VA's outcomes are first collected by the Local Correlation Component of the agent and then forwarded to the client through a Kafka message broker. The VA identifies vulnerabilities in the installed applications based on known CVEs as existing in NIST's National Vulnerability Database (NVD) to produce the results. One of the limitations of this module is that if software running on the platform is not indexed in the National Vulnerability Database (NVD), it can pose a challenge for conducting a comprehensive vulnerability assessment. The NVD is a well-known repository that provides a centralised source of vulnerability information for widely used software and systems. However, it may not cover less popular or custom-built applications that are unique to a specific organisation or industry. However, if the software vulnerability is also based on a misconfiguration from the operating system, the VA module will be able to identify the risk.

### 3.4.7 SIEM System

The SIEM system feeds the interactive Forensics Module with various security-related data across all agent-installer endpoints. It is built on the open-source Wazuh[2], which offers a variety of security-related services that continuously monitor an IT infrastructure. The Wazuh Manager, where data is gathered, processed, indexed, and stored, receives events from lightweight agents that run on the monitored systems and collect all data. As a result, the server level is the only location where security intelligence and data analysis are carried out, ensuring that the resources required at the client level are kept to a minimum. Wazuh clients can be used with a variety of operating systems, such as Windows, Linux, Mac OS X, AIX, Solaris, and HP-UX. The events that the Wazuh agents report is the result of a variety of tasks, including (i) inventory of running processes and installed applications, (ii) collection of log and events data, (iii) monitoring of file and registry key integrity, (iii) monitoring of open ports and network configuration, and (iv) configuration assessment and policy monitoring.

---

[1] https://github.com/rbsec/sslscan
[2] https://wazuh.com/

### 3.4.8 HEIR Machine Learning (ML)-based Anomaly

This module provides efficient event and threat data classification based on specific rules related to cyber security requirements and cyber-threats level of criticality and novel machine- learning (ML) models. In particular, adaptations of existing ML models utilized in anomaly detection are incorporated, which match the requirements of the health systems. The ML module takes the input from HEIR IoT (Logs) and processes the records to differentiate between anomalies and non-anomalies. After that, the ML component processes the results in a detailed report. The result is visualized in the FVT toolkit to represent the results tangibly. The selected model/algorithm based on supervised/ unsupervised learning algorithms depends on each client / hospital (PAGNI, CROYDON). Each use case has its own model which was built to suit the data structure and classify the anomalies in a good order.

### 3.4.9 Local Rama Score Calculator

The Local RAMA Score Calculator is responsible to calculate the Local RAMA score of a hospital or healthcare facilities' connected client. It is also responsible for to construct the metadata related to the identified issues. To calculate the score, the RAMA Score Calculator receives aggregated input from several HEIR Client. To do so, the local RAMA Score Calculator subscribes to the "HeirClientToRama" Kafka topic, receives the aggregated events of the necessary components, calculates the Local RAMA Score, and metadata and continuously provides the output to the 'RamaToHeirGUI', i.e., the topic that HEIR's Aggregator is subscribed.

### 3.4.10 Blockchain-based Privacy-Aware Framework (PAF)

The goal of the PAF is to provide a secure path to a data source, where data access is controlled by a set of policies typically provided by an organization's Governance Officer. The PAF is built on top of the Open Source Fybrik framework [144], which in turn is built on top of leading Open Source technologies such as Kubernetes [145] and Istio [146] for service mesh implementation, and Open Policy Agent [147]. Inside the context of HEIR, the PAF expanded the functionalities of the above technologies to cover a new set of use cases from the healthcare sector.

1. The "Consent" use case tackles the issue of how patient consent to share their records can be incorporated into the PAF mechanism. This use case demonstrates how the FHIR Consent resource within the hospital's FHIR server can be used to hold consent conditions on an individual patient basis. An example can be a patient who grants access of his/her Observation records for research within a given time window. Whereas without the PAF, any researcher with access to the FHIR server can access all Observation records, the policy rules in PAF were defined to return to the HEIR Fybrik module a new data redaction/transformation action called "Consent" if the accessor is a Researcher. The HEIR Fybrik module was extended to accommodate this new action, which checks the consent permission associated with each Observation record to be returned. Records without an associated patient's consent, or which fall outside of the consent time window (i.e., the Observation was performed outside of the time window) have all PII information redacted before being returned. Records which have the correct consent permissions and timestamp are returned unredacted.
2. A second use case arose from discussions with a clinician from the Norwegian Diabetes Registry team. There are many healthcare registries distributed across Norway, however each registry is individually managed. This means that a clinician or researcher who wants to cross-correlate data between registries (for example, the Diabetes registry with the Prescription registry) needs to fill out many request forms and wait to obtain access approval – a process that typically takes upwards of a year. HEIR tackled this by prototyping a solution where the PAF transparently federates the distributed registry data stores and issues queries for data across all the linked data stores. PAF then applies its policy-driven data transformation/redaction actions across the returned federated data set. Our prototype was directed at two main actors – Patients and Researchers. Patients can obtain their records from all of the linked registries, essentially combining these formerly siloed data stores into an integrated electronic healthcare record. Researchers can request records from across this federated linkage, however PAF policy causes PII information in the returned records to first be anonymized by use of a hashing algorithm. As in our other use cases, users identify, and role is embedded in a cryptographic token (JSON Web Token) which must accompany every request for data.

### 3.4.11 HEIR Forensics Visualization Toolkit (Interactive Forensics)

The Interactive Forensics module is based on AEGIS' FVT [148] to display analytical information about the captured security events and relevant system information. The module allows users to drill down to individual assets monitored by HEIR and see monitoring metrics as well as events captured by the SIEM in order to gain situational awareness in a short time. The FVT provides users with a timeline-based representation of the security events captured by the SIEM sub-module and processed by ML Anomaly Detection Module. It is accessed through the 1st layer of visualizations and is meant to represent the captured events in a more detailed way.

#### 3.4.11.1.1 Overview and Devices

Users accessing the FVT will firstly see the 'Overview and Devices' screen of the selected client (Department). Generic information about the selected client (e.g., connected devices, total critical events detected, security status etc.) and the calculated Local RAMA score are presented in the top panel. (Figure 7)

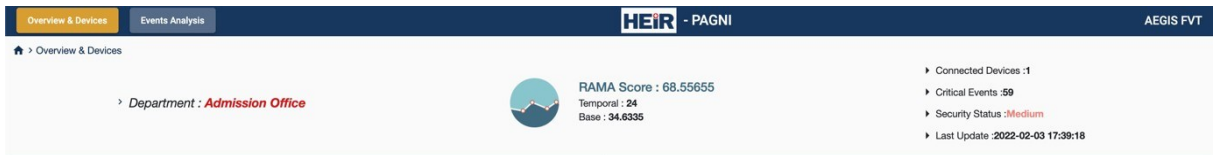

*Figure 7: RAMA Score*

The results of the HEIR Exploit Tester and Cryptographic Checker are displayed via graphical representations and lists (Figure 8).

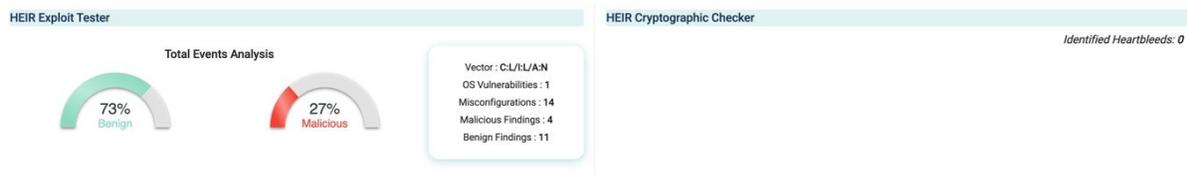

*Figure 8: HEIR Exploit Tester and Cryptographic Checker*

The Vulnerability Assessment's metadata are demonstrated in an expandable section, where the users can quickly identify the top 10 vulnerabilities and the full list of vulnerabilities per application. Each vulnerability is clickable and linked to MITRE's CVE [149]. (Figure 9)

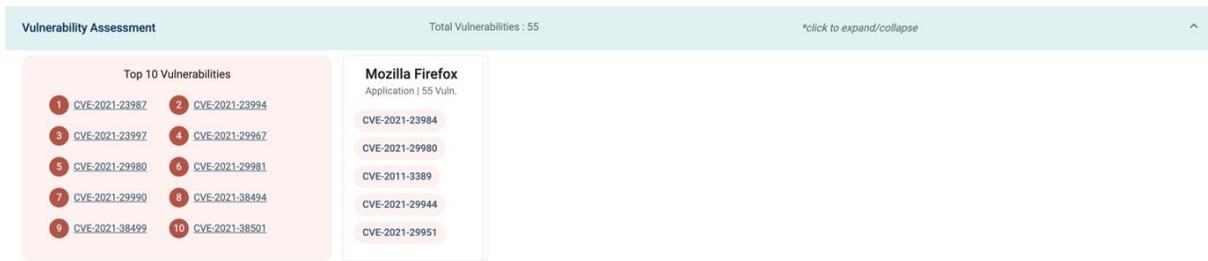

*Figure 9: Vulnerability Assessment*

HEIR Network Module's analysis results are currently presented in a tabular view. Critical information, such us the number of exploits, attacks etc., are presented in the top of the section. In addition to sorting capabilities, there is also a standalone text filter, so the users can search through the available information and quickly identify meaningful details (Figure 10).

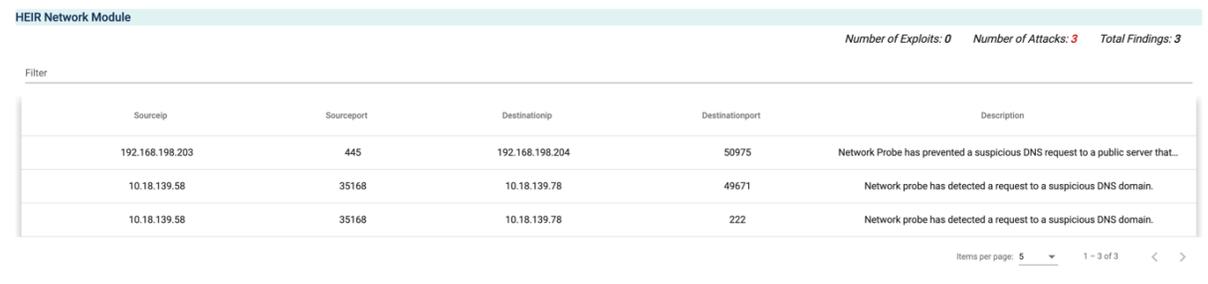

*Figure 10: HEIR Network Module*

At the bottom of the screen, users can choose from the devices (that are connected to the HEIR Client) the one they want to investigate. (Figure 11)

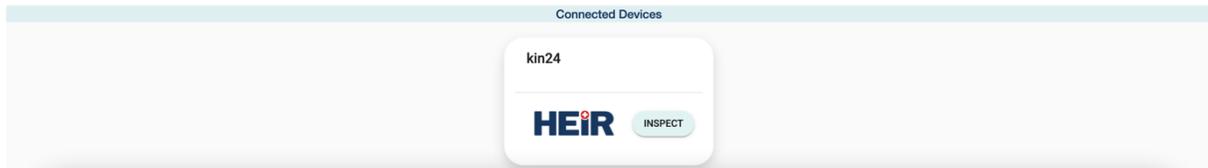

*Figure 11: Connected Devices*

The complete page of HEIR Client's Overview and Devices is available in Figure 12 below.

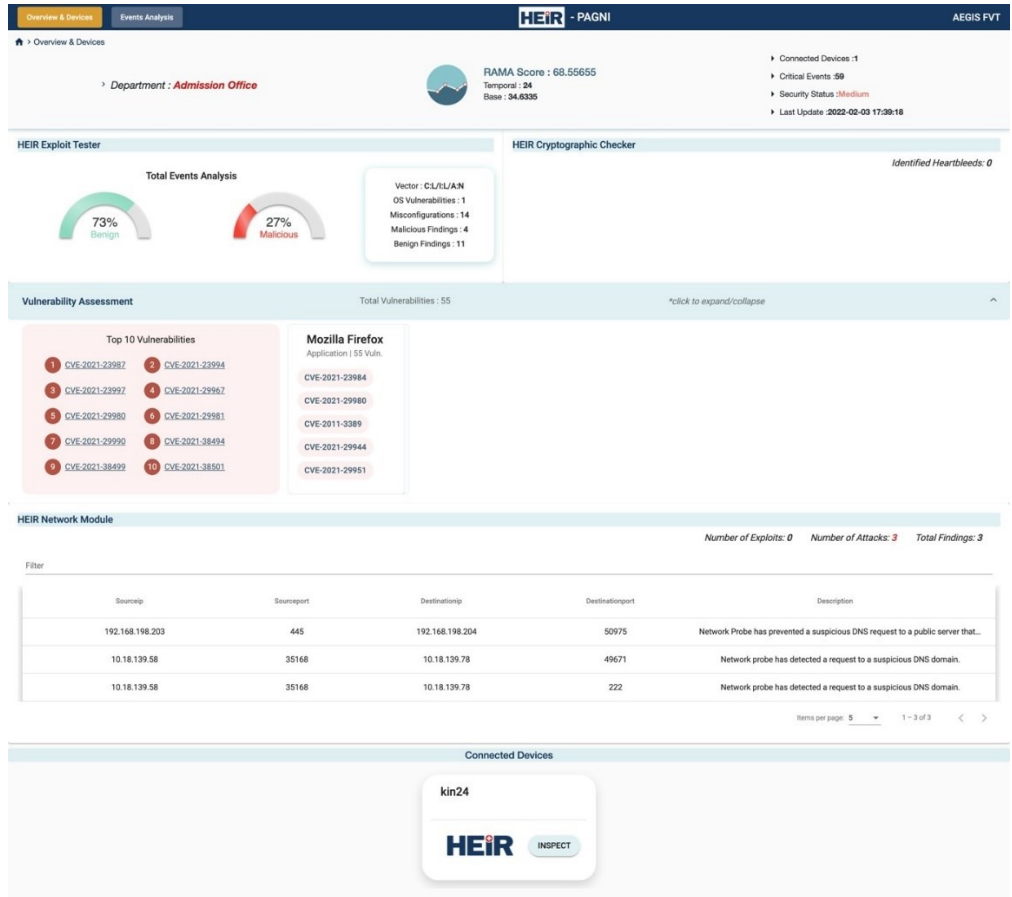

*Figure 12: HEIR Client's Overview and Devices*

3.4.11.1.2  Inspect device.

The FVT's device inspection main dashboard (Figure 13) allows users to choose from a set of widgets containing different types of visualizations, that refer to different System metrics and Security event related information. The widgets could be standalone and support discrete input sources, but their combination offers a comprehensive depiction and meaningful visualization of the data. The user will also be able to filter the incoming data by different meaningful parameters and to search historical data.

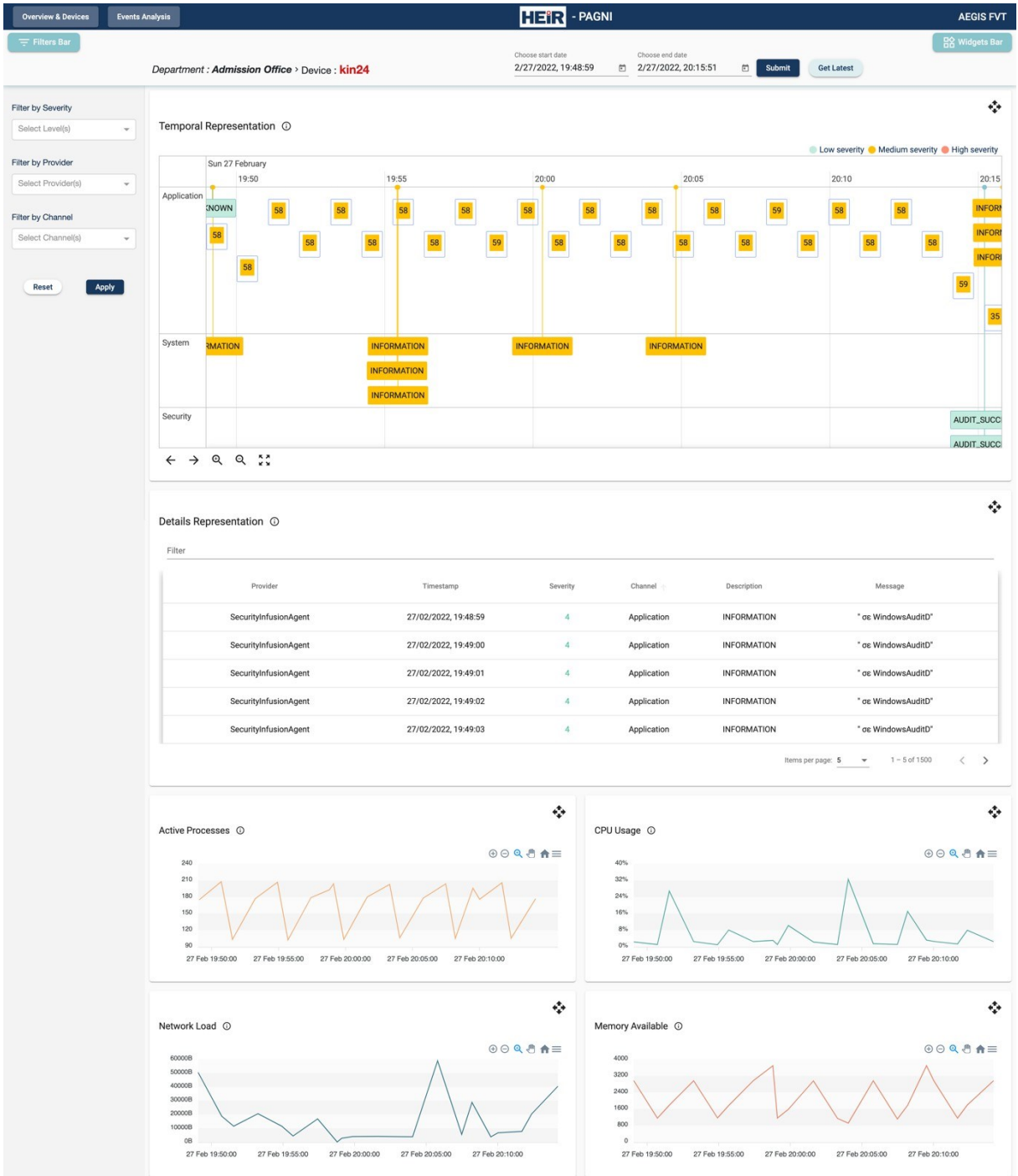

*Figure 13: Inspect Device page*

Temporal (point in Time) representation of incoming logged events captured by the SIEM will be available through the Timeline widget (Figure 14). By changing the date period ('Start date' to 'End date') in the timeline all the available widgets will be timewise synchronized and updated accordingly. In the filter bar (widget menu in top left side), the user can further filter the results by Severity, by data provider (e.g. Security Infusion Agent, control manager etc.) and by channel (e.g. application, system etc.).

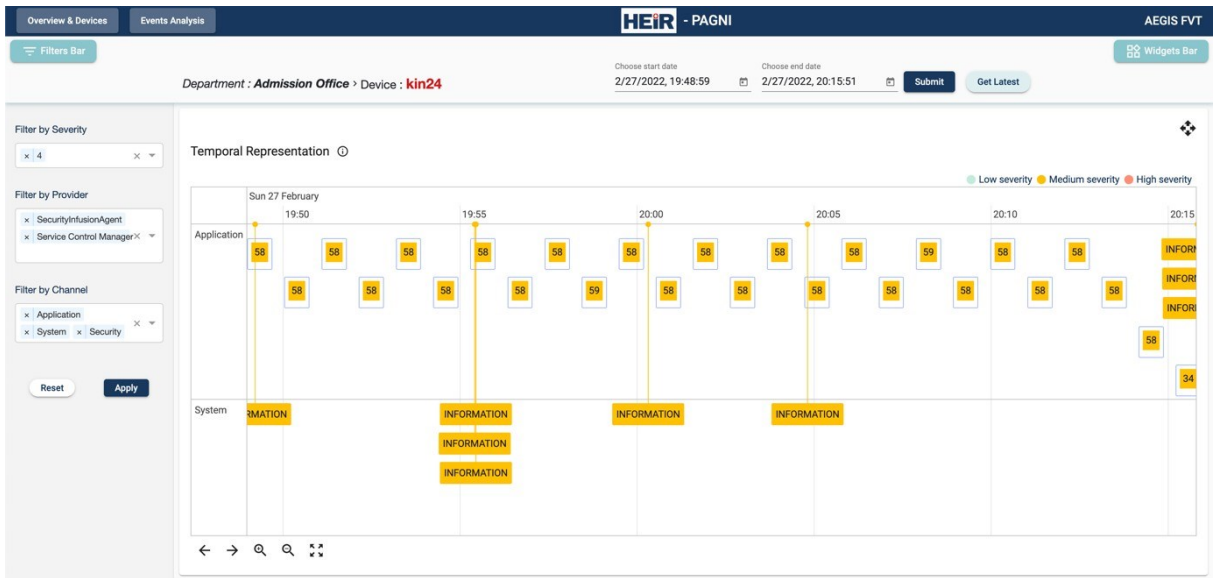

*Figure 14: Temporal Representation*

Detailed information about the incoming events will be presented in the Details widget (Figure 15). This widget also supports a standalone text filtering capability in order to search through the available information.

*Figure 15: Details Representation*

A variety of different device-related metrics can also be analyzed through the available Line Chart widgets (Figure 16), including the active processes, the CPU usage, the network load and the available memory of this device. The corresponding widgets support zooming, panning, downloading options and are interconnected as they use the same time series (x axis).

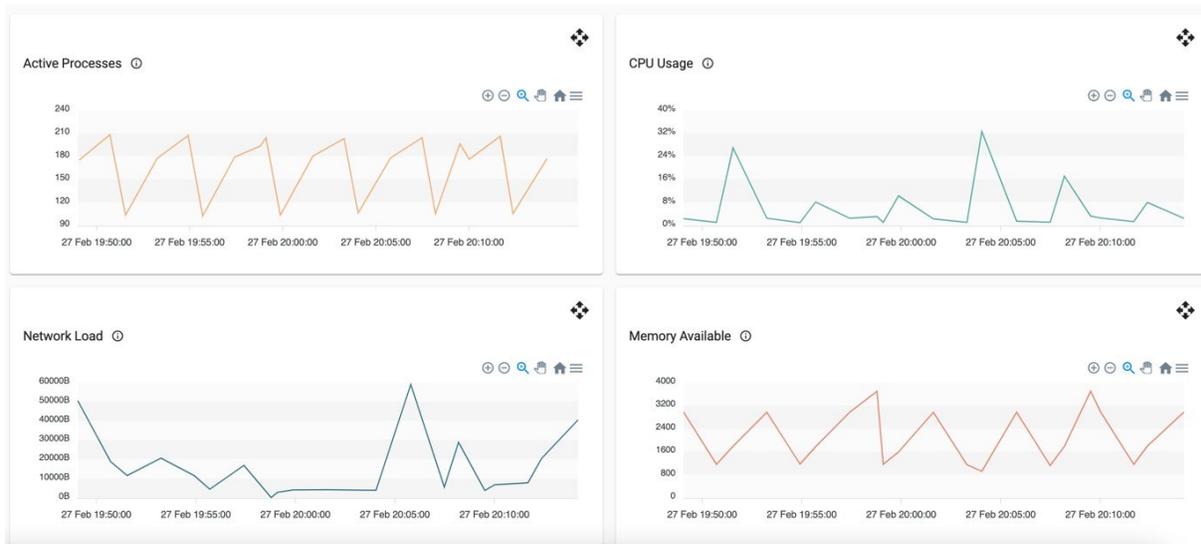

*Figure 16: Device's Metrics*

#### 3.4.11.1.3 Events Analysis

FVT also provides an Events Analysis screen, in which the logged events from the department's (client) connected devices and the output of the Anomaly Detection's module (ML) are presented. The top layer of the screen contains the current available filters and the datetime picker for historical data requests (Figure 17).

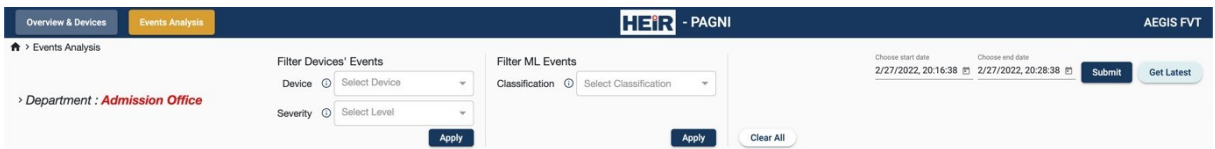

*Figure 17: Events Analysis filter section*

Temporal (point in Time) representation of incoming logged events captured both by the Anomaly Detection Module (ML) and the selected devices will be available through the Timeline widget (Figure 18). By changing the range period in the timeline all the available widgets will be timewise synchronized and updated accordingly. At any time, the current zoomed period is displayed above the Timeline (upper right corner). A functionality to select a connected device and navigate to FVT's device inspection main dashboard (Figure 18) in the current zoomed period, is available. This functionality aims to enhance forensics analysis, by offering timewise parallel comparison of the detected anomalies in the department and the devices' logged events.

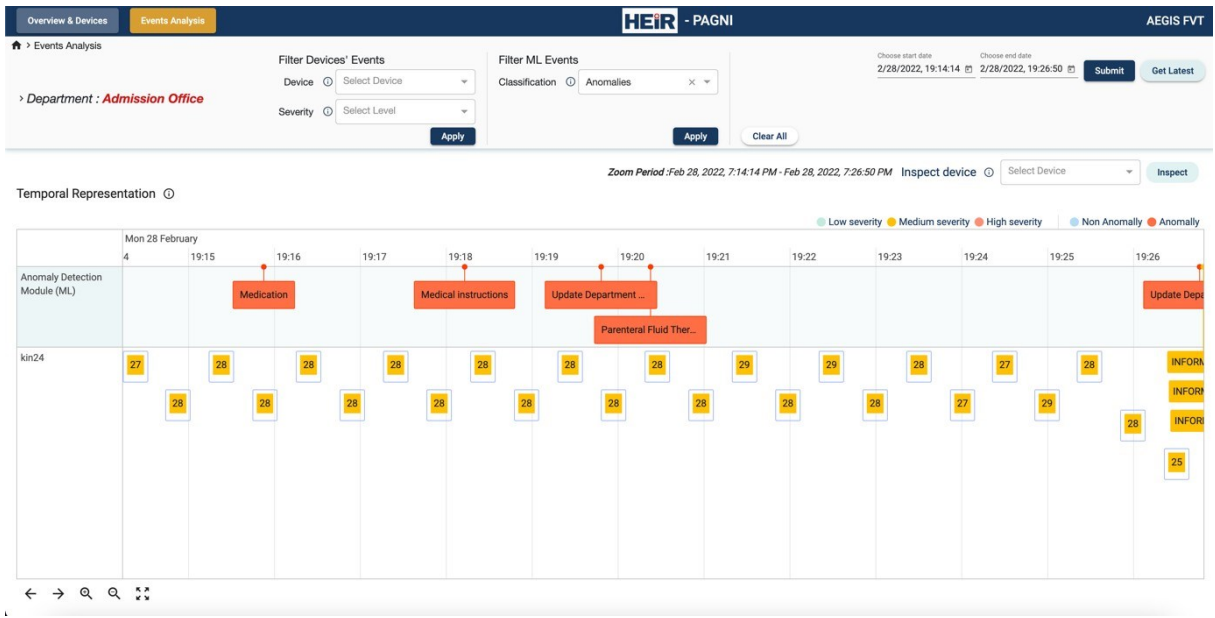

*Figure 18: Temporal Representation*

Detail representation widgets, both for the Anomaly Detection Module (ML) output and the logged events (SIEM), are available (Figure 19). Independent filtering and sorting capabilities are also available for these widgets (e.g. sort by a specific column, or focus on specific table line).

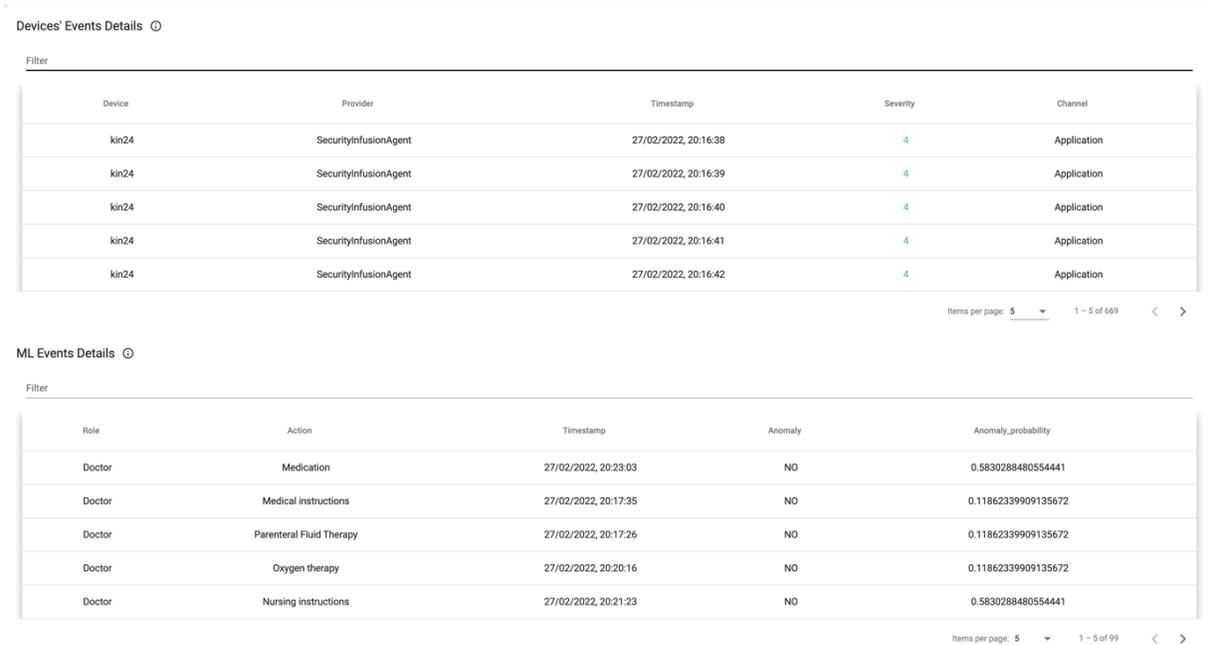

*Figure 19: ML Events and Devices' Events Details Representations*

The full page of Events Analysis is available below, in Figure 20.

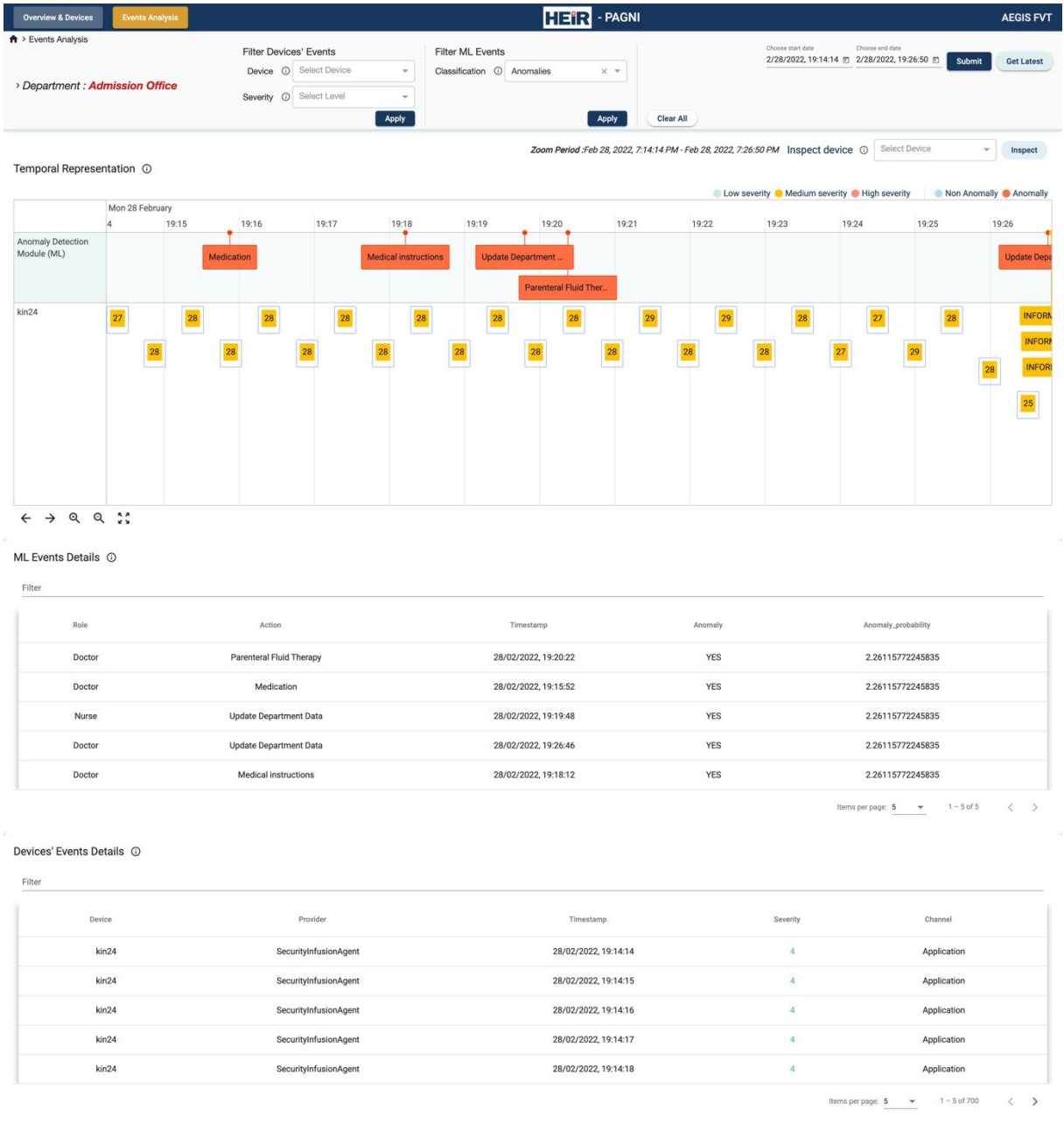

*Figure 20: Events Analysis Full Page*

The main screen of FVT is presented in Figure 21.

### 3.4.12 HEIR 1st Layer GUI (1st Layer of Visualization)

HEIR's 1st Layer GUI includes visualizations of information generated by the 1st level services running inside the hospital environment. This component runs as a containerized service which fetches data found in the local storage facility. The displayed data includes RAMA Score and associated metadata generated by the HEIR Client. It is also linked with the Forensics Visualisation Toolkit (FVT). Figure 22 depicts the main screen of the GUI. Users can drill down to further security information by clicking Inspect Client, which navigates them to the HEIR Interactive Forensics GUI.

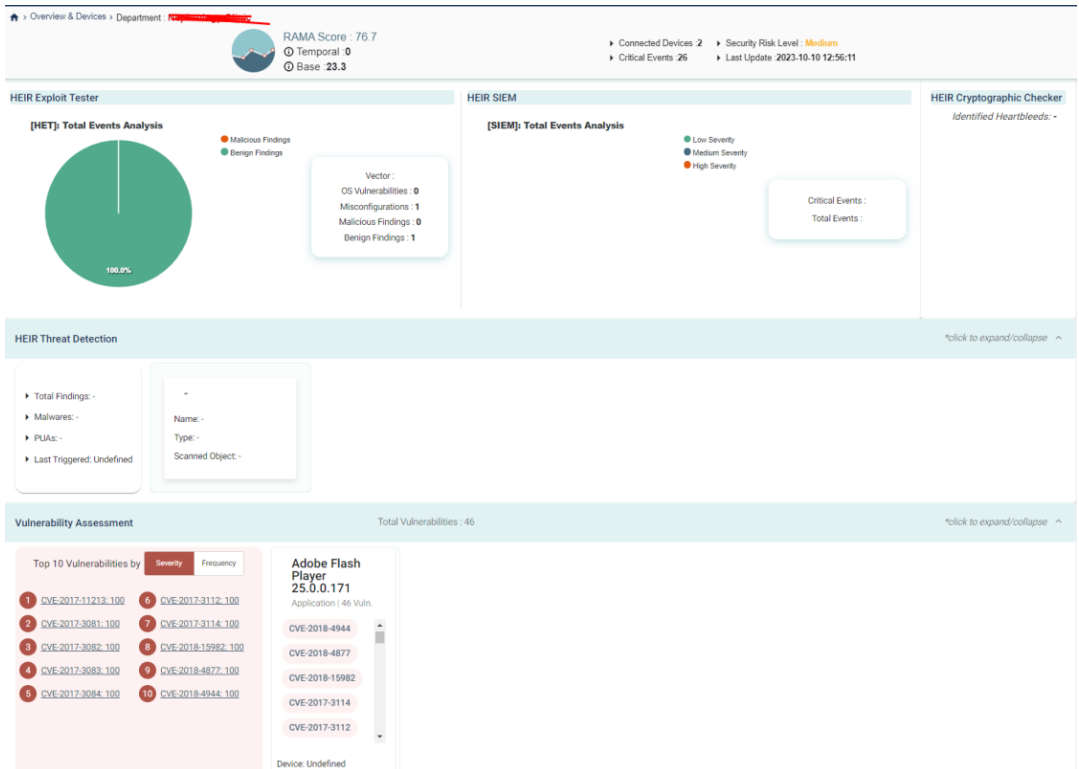

*Figure 21: FVT Main page*

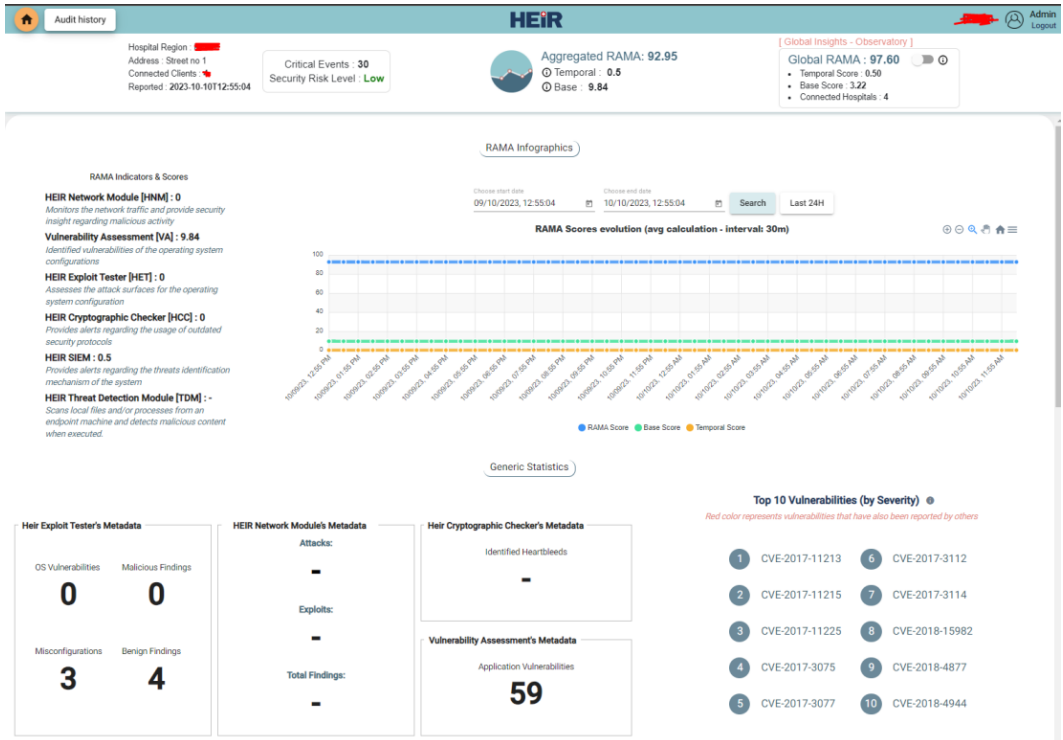

*Figure 22: HEIR's 1st Layer GUI – Main Page*

### 3.4.13 HEIR Aggregator

The HEIR Aggregator is a component designed for health institutions running multiple independent departments for which an aggregated cybersecurity status is required. The Aggregator is responsible to calculate the Aggregated Local RAMA score for a healthcare organization, as well as anonymise private information before transmitting the score and the metadata to the Global RAMA score calculator. An aggregated local RAMA score is also computed after having been provided with multiple local RAMA scores by the HEIR clients deployed on the individual departments. The HEIR Aggregator input is given by JSON files written by the Kafka connector to an Elasticsearch storage in the "rama-heir-gui" index. The HEIR Aggregator is triggered based on a user-defined schedule (e.g., hourly), read the most recent outputs from the HEIR clients in the Elasticsearch storage, compute the aggregates, and write the aggregated values for RAMA and the event statistics to the Elasticsearch storage, where they can be accessed by the HEIR GUI. The Aggregator finally sends its output to the HEIR Observatory. The process of compiling the cybersecurity status from the HEIR clients deployed in one health institution is shown below in Figure 23.

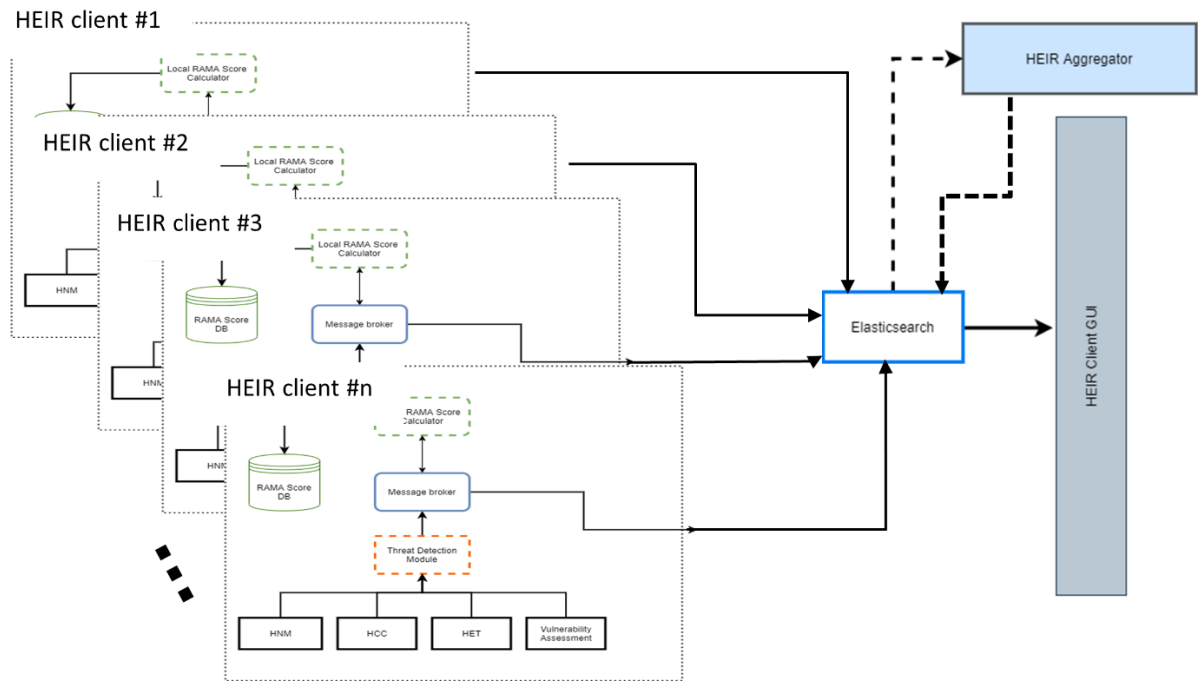

*Figure 23: HEIR Aggregator*

### 3.4.14 HEIR global benchmarks (Global RAMA Score Calculator & The Security and Privacy Assurance (SPA) suite)

The HEIR global benchmarks are composed of two components, the Global RAMA Score Calculator and SPHYNX's Security and Privacy Assurance (SPA) suite. The former enables healthcare stakeholders to identify common issues for different healthcare sectors, by calculating the Global RAMA score. It receives input from the HEIR aggregator and acts as an aggregator of all the local RAMA scores of a healthcare facility providing a unified score. As for the latter, the penetration testing component of the suite was utilized to further strengthen the cybersecurity status of the involved pilots by executing penetration tests and providing reports to the pilot owners.

### 3.4.15 Observatory Database

The HEIR Database stores the data sent by the deployed HEIR Clients. This data includes RAMA scores and other generated outputs that are relevant for the parameters of the experimentation protocol.

### 3.4.16 Analytics Engine

The Analytics Engine is responsible for collecting and analyzing the data from the Observatory DB in order to provide global insights, statistics, and recommendations. Moreover, the Analytics Engine supports the historical analysis, advanced queries mechanisms, and user interaction capabilities, based on the role and the requirements of the end-user.

### 3.4.17 HEIR Observatory (2nd layer of Visualization)

The HEIR Observatory is responsible to collect, analyse and present the results, in an anomyised form, of all the hospitals involved in the HEIR ecosystem, in order to provide global insights on the level of security in healthcare environments. The main page of the Observatory displays statistics of the Global RAMA Score and the historical evolution of data like the captured events or detected vulnerabilities. Figure 24 below illustrates the main page of the Observatory.

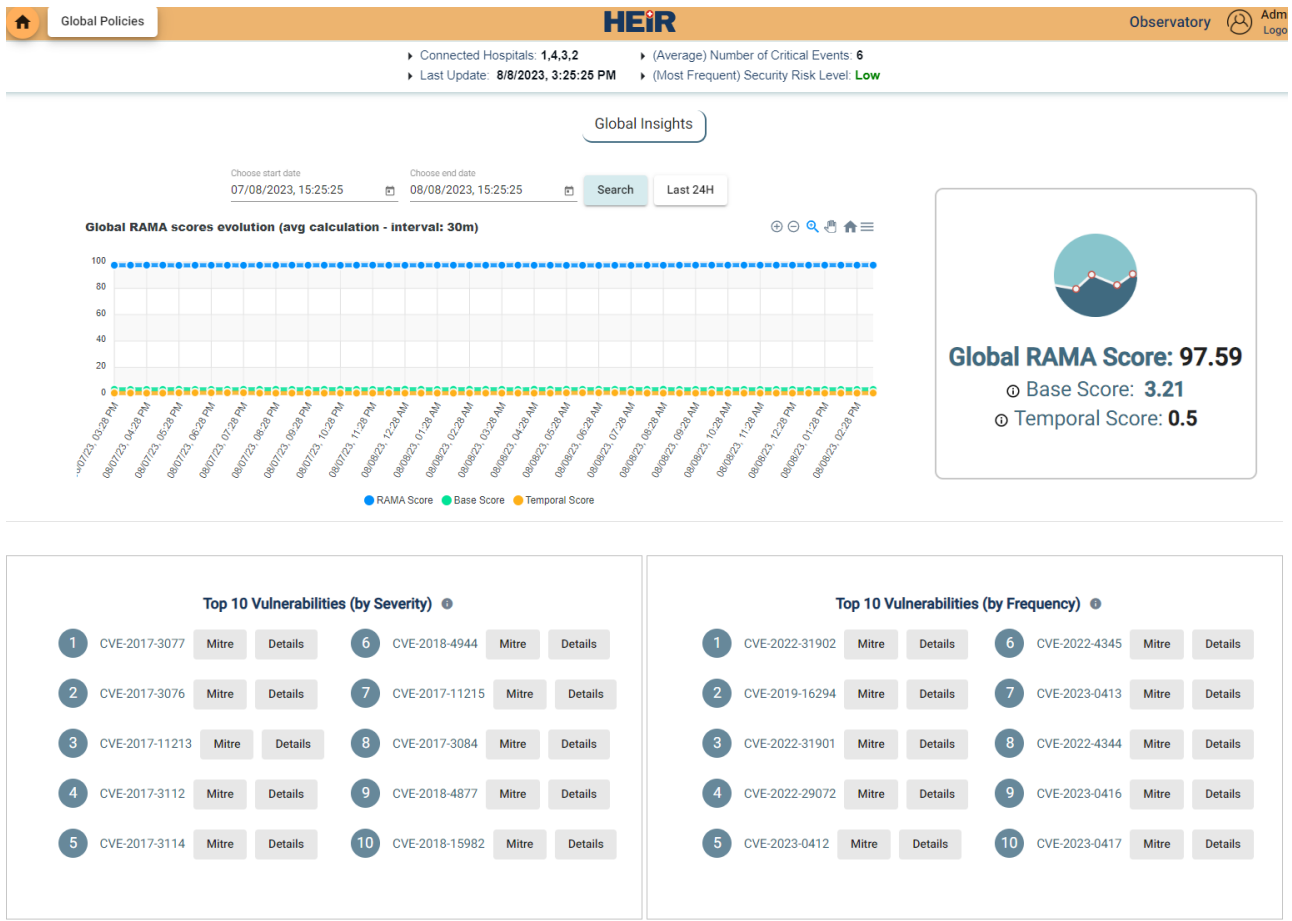

*Figure 24: Observatory Main page*

# 4 Validation

Four healthcare providers (end users), HYGEIA Hospital (HYGEIA), the General University Hospital of Heraklion (PAGNI), the Norwegian Centre for E-Health Research (NSE) in cooperation with the Norwegian Diabatesregister for Adults (NOKLUS) and the Croydon University Hospital (CUH) were selected to demonstrate the HEIR framework on four diverse, real- life health pilots in sensitive medical environments.

## 4.1 Pilot sites and assets to protect.

### 4.1.1 PAGNI

PAGNI's use case focused on the Hospital Information System (HIS), which allows healthcare providers (physicians, nurses, etc.) to access patient data stored in the system during their regular activity in the clinic. The objective was to protect the "PANAKIA" HIS system from cyberattacks. Healthcare providers can register patient information, request or view exams from the Laboratory Information System and the Picture Archiving and Communication System, and access data from the Surgery List through this platform. The primary intention was to secure the system by either protecting the workstation or utilizing machine learning techniques to detect anomalies in the platform's daily use by medical personnel.

### 4.1.2 HYGEIA

HYGEIA used the HEIR Framework to enhance/supplement the security and data privacy of its IT architecture supporting the "***my- Ygeia***" [150] mobile application, that the patients use to access, manage, track, and share data, contained in their Personal Health Records (PHRs). "*my-Ygeia*" is a relatively recent application that is constantly being enriched with new features that will allow patients to access their health data, and HYGEIA to offer extended homecare services. The app is already used by tens of thousands of users; thus, the maximum assurance of security and data integrity are imperative. The demonstration took place in the test environment that accurately replicates the sub-section of the IT infrastructure responsible for the operation of the *"my-Ygeia"* application. For compliance purposes, the data therein were test/pseudonymized data that conform to regulatory requirements. It was deemed as the most appropriate but also useful use case, for HYGEIA to demonstrate the use of the HEIR Framework for the proactive management of system vulnerabilities that can compromise the IT infrastructure.

### 4.1.3 NSE and NOKLUS

The overall aim of this use case was to make the exchange of data between diabetes patients and databases/registries more secure for the parties involved. Initially, the so-called "Diabetes Diary App" [151] was to be used for gathering the required patient data. This application allows diabetics to record – among other values and functionalities – glucose values, insulin doses and carbohydrate intake. However, it became apparent in the early stages of the project that this application would not be the optimal choice, as the data it contained was self-reported and had to be maintained manually by its users. Therefore, an alternative had to be found, which was identified in cooperation with the Finnish company Sensotrend [152] which is easier to use and its processed data was stored in the HL7 FHIR standard. This simplified the communication and transmission of data in this project's context. Once these baseline conditions were established – the use of the devices to gather the relevant data as well as the Finnish platform – NSE and Noklus were actively involved in the design of the use cases. Regarding the transfer of the collected data to NOKLUS; the choice fell on the so-called *"DIPS Communicator"* [153], a tool that is used by multiple actors in the healthcare sector to communicate patient-sensitive information. DIPS Communicator is among the most widespread communication solution in the health sector in Norway. The decision was made, since the DIPS Communicator was at this stage already used by NOKLUS and thus integrated in their system. In close cooperation with the University in Tromsø (UiT), it was agreed to house the individual components in Microsoft's Azure, while the necessary licenses were also provided by the UiT. After the technical framework conditions within the use case had been clarified, it was agreed within the overall project that NSE and NOKLUS should implement and demonstrate the functionalities of the PAF – a secure data transfer from the patient to various recipients in their use case.

### 4.1.4  CUH

The asset to secure in this case was a medical application system, that can be hosted in a separate stand-alone virtual machine, upon which the HEIR-trial software could be installed to assess its functionality. A selection of medical applications that were used at Croydon at that time was therefore assessed, and the selection was the "TEAM 3 device, a Fetal/maternal monitor" [154] which has multiple modalities such as – heart rates of fetus and mother, pulse oximetry, blood pressure, as well as uterine muscle activity – uterine contractions (i.e., sound, light, electrical and mechanical means that needed to be recorded.) The output of the device was routed to a central Sonicaid server, and the entire system could be duplicated. Furthermore, there was already a live database that was available for review as the benchmark. A VM machine and laptops were secured, with the inclusion of a Team 3 device and the sonicaid server networked to provide a secure repository for the HEIR agent to be installed. It also replicated the current working environment found within the labour ward. Additionally, an agreement with the third-party vendor for assistance and support as well as permission to use their device in such a manner was also possible.

## 4.2  Execution of real-life demonstrations

Playbooks are generally intended to specify actions according to predefined framework conditions. With the help of these playbooks, the following demonstrations of the pilot partners are to be presented in a coherent and easily understandable way.

### 4.2.1  PAGNI

PAGNI aims to demonstrate the function of the Threat (Anomaly) Detection components of the HEIR Project, which focuses on the Hospital Information System (HIS) and access to the patient data held therein. Healthcare providers (physicians, nurses etc.) have access to the HIS during their daily practice in the clinic. From that system, called PANAKIA, the Healthcare providers can record information about the patient, request, or view exams from the Laboratory Information System (LIS) and the Picture Archiving and Communication System (PACS) and retrieve data from the Surgery List.

#### *4.2.1.1  Data gathering and collection.*

The HEIR 1st layer GUI was deployed in the premises of PAGNI at the dedicated environment for HEIR. HEIR's Threat Hunting Module collected, analyzed and correlated the results of all tests run by the HEIR Client in any device or system, facilitating the work of IT professionals in medical environments and displayed their current security status in terms of adaptation of good practices. Main components of the HEIR Threat Hunting Module are the Novel HEIR Client, the Local RAMA Score Calculator, the 1st Layer GUI and the HEIR Aggregator. Figure 25 presents the HEIR 1st Layer GUI at the PAGNI environment.

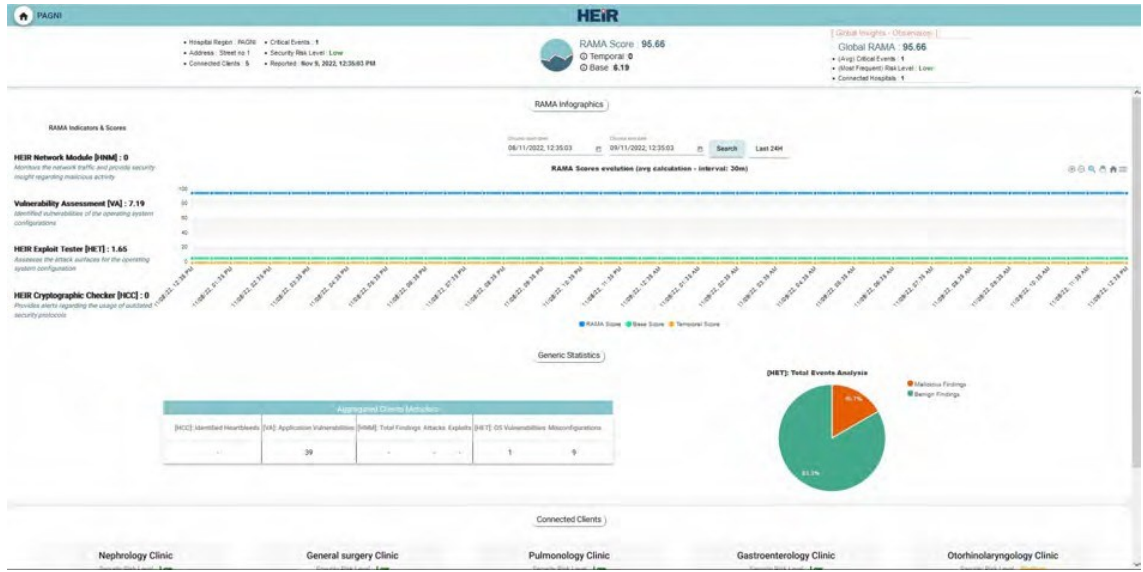
*Figure 25: HEIR 1st Layer GUI at PAGNI*

### 4.2.1.2 Infrastructure and Architecture

The PAGNI HEIR environment consisted of two dedicated Servers 2 Desktop VMs, Virtual Machines (with the same configuration as the existing PCs in the Hospital), along with four productive workstations from different departments of the hospital that the HEIR components are deployed. The main actors were the IT SysAdmins (System Administrators) of the hospital. The IT SysAdmins managed the Roles, Rights and the Users of the system. The system was also accessible from the Primary Care Units in the region of Heraklion (outside of the hospital) using a VPN connection. The overall architecture is presented in Figure 26.

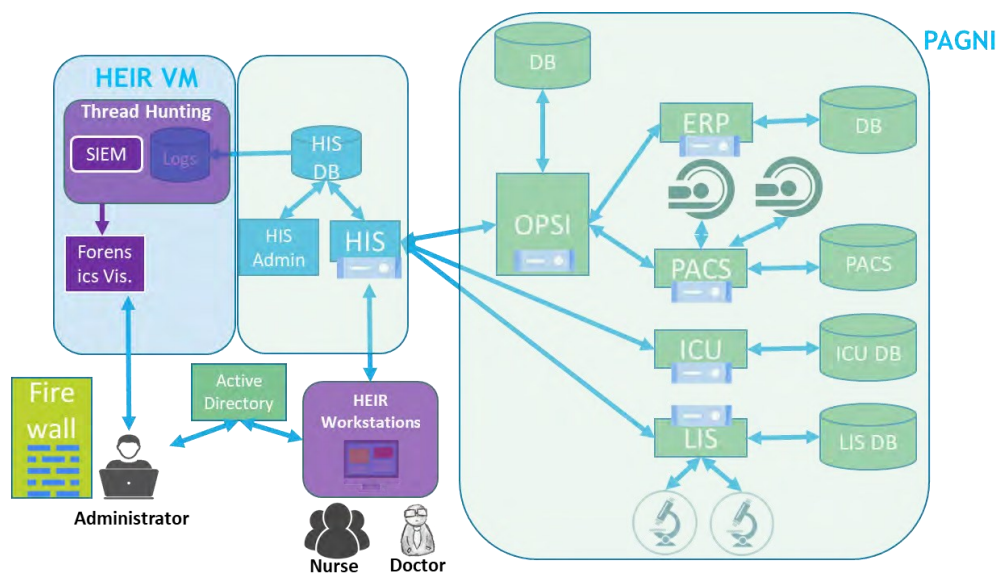
*Figure 26: PAGNI Use Case – Overall Architecture*

### 4.2.1.3 Playbook Scenarios

The following section explains the three scenarios foreseen in the playbook, namely the Identification of outdated software, Threat hunting using HEIR' Threat Detection Module and SIEM system, Threat detection using HEIR'S Threat Detection Module and SIEM system and the Machine learning-based Anomaly Detection.

4.2.1.3.1   Identification of outdated software - Involved components: HEIR's Vulnerability Assessment module, HEIR's 1st Layer GUI, HEIR's Local RAMA

In this scenario, the administrator of the hospital uses the HEIR's 1st Layer GUI to check whether any outdated software exists within the organization. More specifically, the administrator first checks the local RAMA score and identifies that one of the connected clients (a workstation at a clinic) has reported vulnerabilities. The administrator then proceeds to further examine the metadata for the specific client, as reported through HEIR's Vulnerability Assessment. The reported metadata informs the user that a client has an outdated version of the Mozilla Firefox browser.

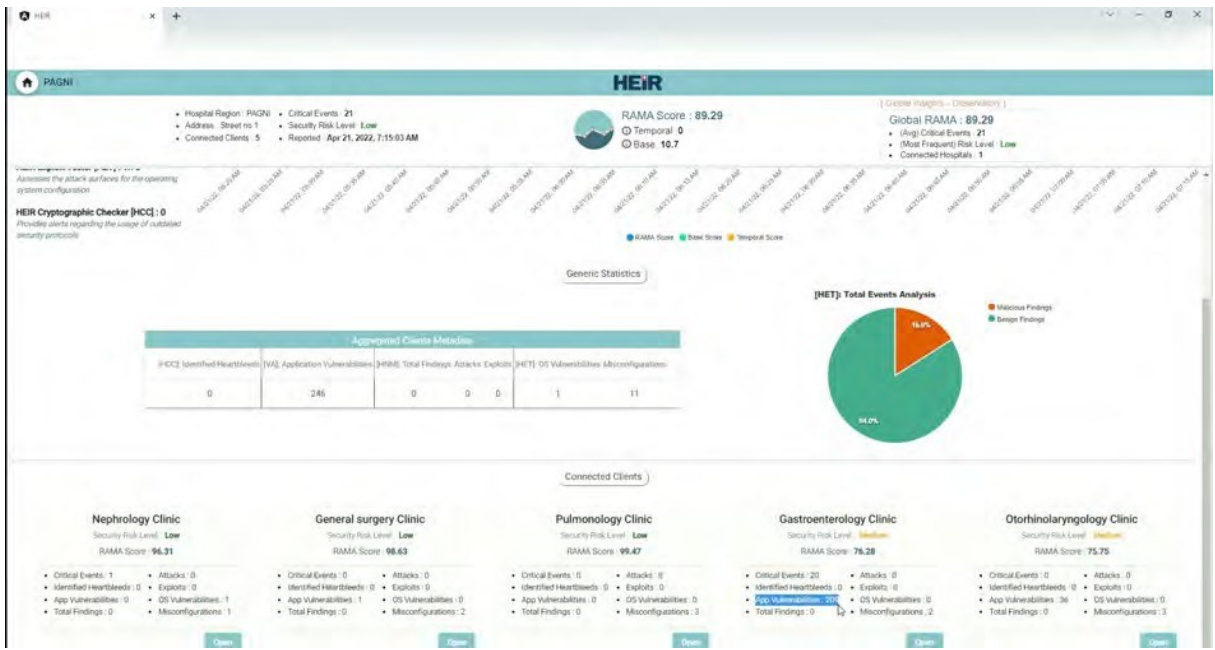

*Figure 27: Outdated software detected (1)*

The administrator updates the software (Mozilla) to the specific workstation using remote connection or with physical access at the department/clinic. Then, back at the office, opens again the 1st Layer GUI of HEIR and identifies that there are no vulnerabilities anymore (Figure 29) for the specific client and that the local RAMA score has been increased.

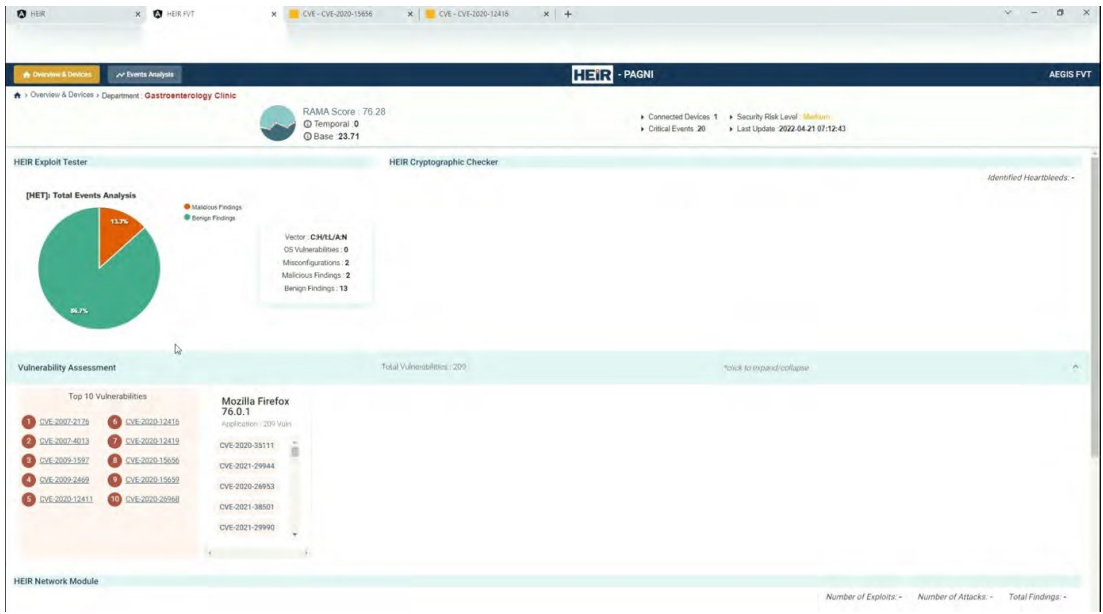

*Figure 28: Outdated software detected (2)*

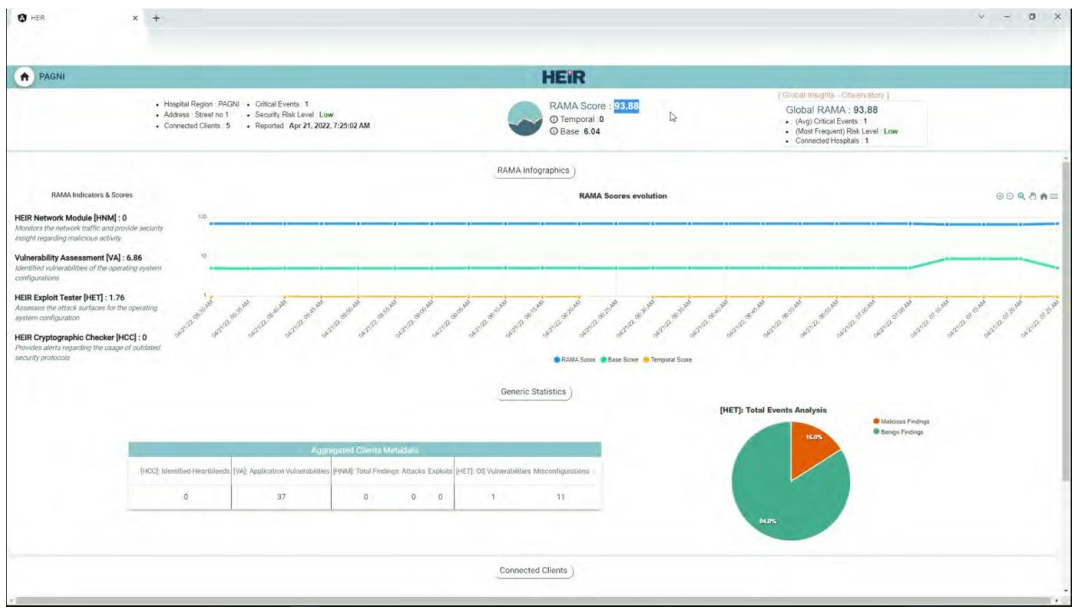

*Figure 29: Increased RAMA score after software update*

4.2.1.3.2   Threat hunting using HEIR' Threat Detection Module and SIEM system - Involved components: HEIR's Network module, HEIR's Threat Detection Module, HEIR's SIEM system.

For the smooth operation of the hospital IT systems, a threat detection and mitigation system that acts quickly and efficiently is crucial. HEIR security mechanisms can detect threats to the servers and workstations that belong to the hospital, neutralize them and inform the IT department for the issues and the actions needed. In this scenario, an end user of the hospital e.g., a physician uses the workstation at the office and opens an email. The email - from a patient - contains a URL and the sender claims that this is an MRI exam for evaluation. The physician opens the URL that is actually a malicious URL. The threat detection module of HEIR identifies the threat and reports to the HEIR platform. At the same time the administrator of PAGNI sees at the 1st Layer GUI of HEIR a threat detection. Checks the local RAMA score and identifies that one of the connected clients (a workstation at a clinic) has malicious findings as shown in Figure 30.

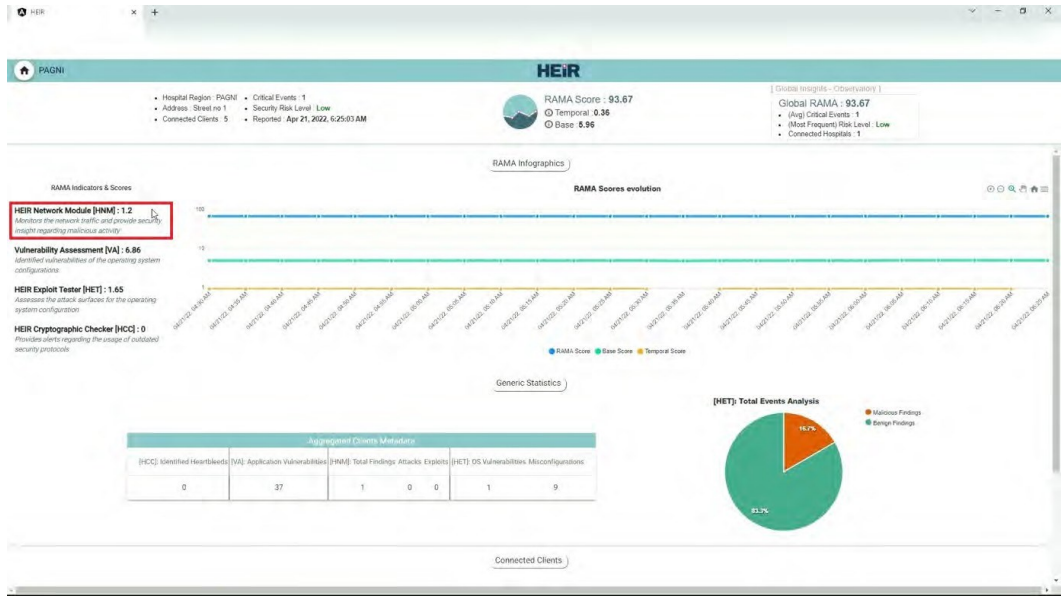

*Figure 30: Local RAMA score reduced*

The administrator opens the vulnerabilities details for the specific client and the system informs the administrator that the HEIR threat detection module identified a malicious network traffic as shown in Figure 31.

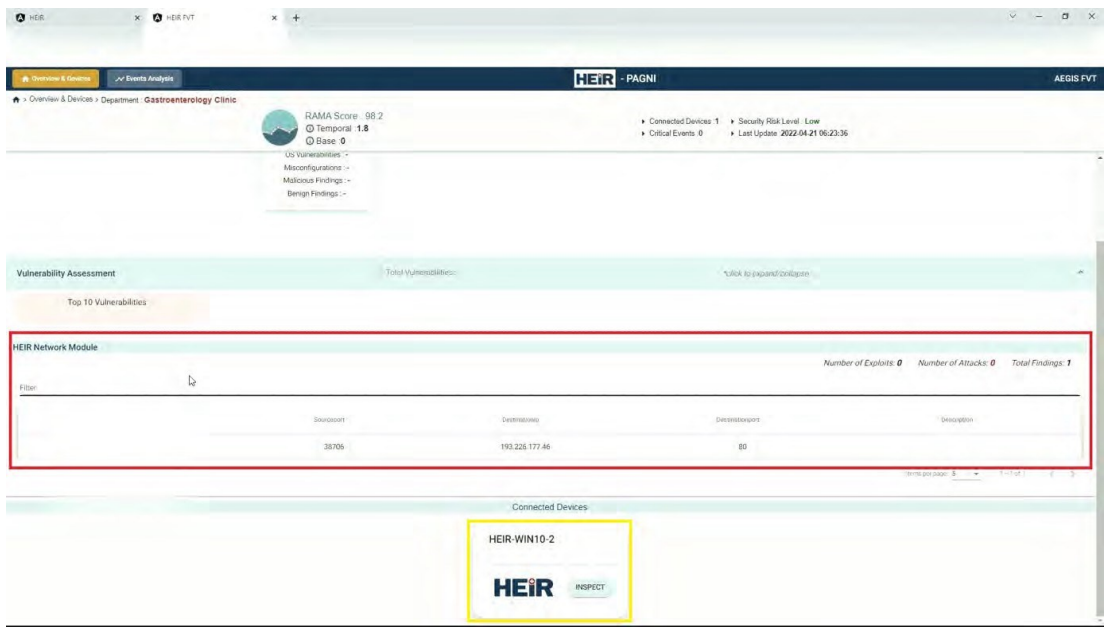

*Figure 31: Malicious network traffic detected*

The administrator subsequently uses the specific workstation and runs a threat scan and mitigation efforts are enacted to properly neutralize the threat, opens again the 1st Layer GUI of HEIR and identifies that there are no vulnerabilities any more for the specific client and that the local RAMA score has been increased as shown in Figure 32.

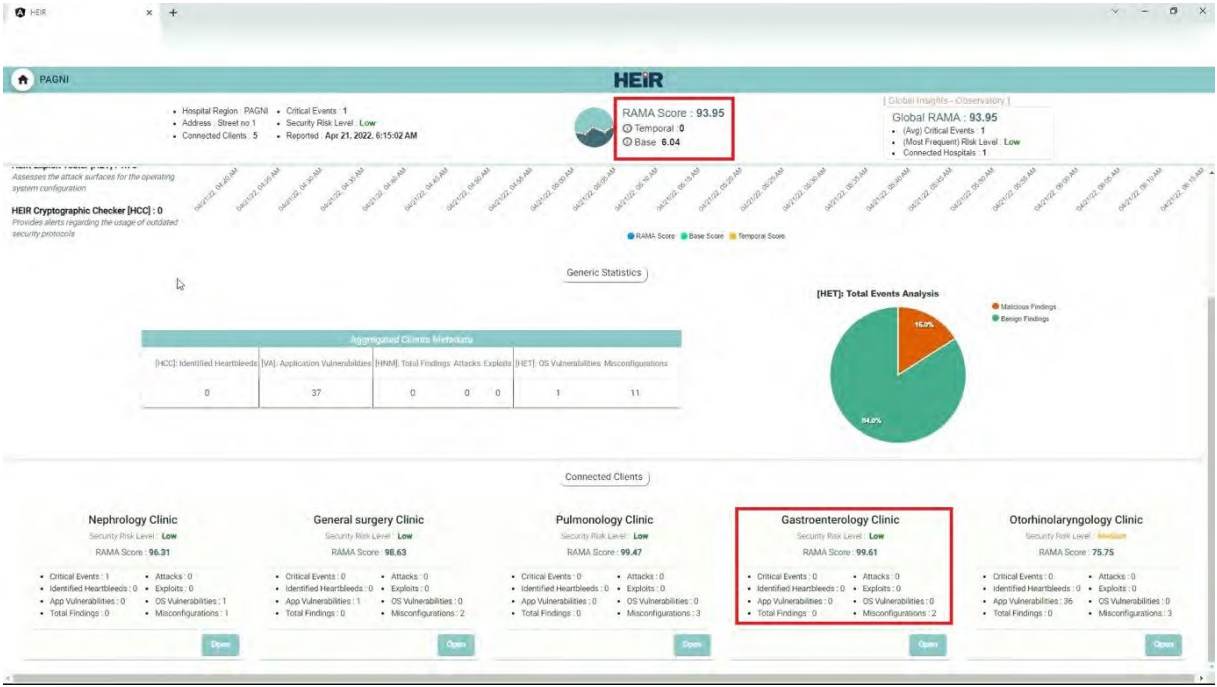

Figure 32: Increased local RAMA (good condition)

4.2.1.3.3 Threat detection using HEIR'S Threat Detection Module and SIEM system - Involved components: HEIR's Threat Detection Module, HEIR's SIEM system, HEIR's RAMA

This scenario involves an external storage device and a malware that attempts to escalate privileges and try a lateral movement tactic. In this case the malicious application is copied from the external storage and then executed unwittingly by the end-user. The threat detection module can detect the malware and reports it to the HEIR platform. The SIEM component is also able to detect critical events performed by an entity, be it user, attacker, or process. In this case the malware - after its execution - modifies the user groups to obtain some privilege escalation. Additionally, it tries several predefined logins to obtain a lateral movement. As the logins do not work, they will generate failed login events that will be considered as an indicator of a compromise, for a SIEM point of view. This type of events, depending on their severity will be submitted to the HEIR platform to be visible and the RAMA score will be decreased. The SIEM and Threat Detection Module work in a complementary way. Let us presume that the detection for the malware was not available, the SIEM high severity events would be reported providing a clear indication that an attack or malicious action are active in the indicated endpoint. On the other hand, the reports from Threat Detection Module to the SIEM can also increase the importance of maybe not important action but they may provide context into the administrator analysis.

4.2.1.3.4 Machine learning-based Anomaly Detection - Involved components: HEIR's ML-based Anomaly Detection Module, HEIR's 1st Layer GUI

Retrospective and anonymized log data are available for the training of machine learning algorithms developed by the technical partners of HEIR. The main components of the IT infrastructure of the hospital (Hospital Information System, Laboratory Information System and Picture Archiving and Communication System) fed the HEIR environment with about 2.5 million records (logs/actions) anonymized logs that cover a period of 20 months. The records contain information about the user, the case/admission the action

(e.g. medical action, medical report, medical history (update), exit report, document upload, surgery report, physiotherapy, request lab exams, etc.) and the connection type (VPN or not). The aim is to identify not typical behavior of the users and alert the IT department.

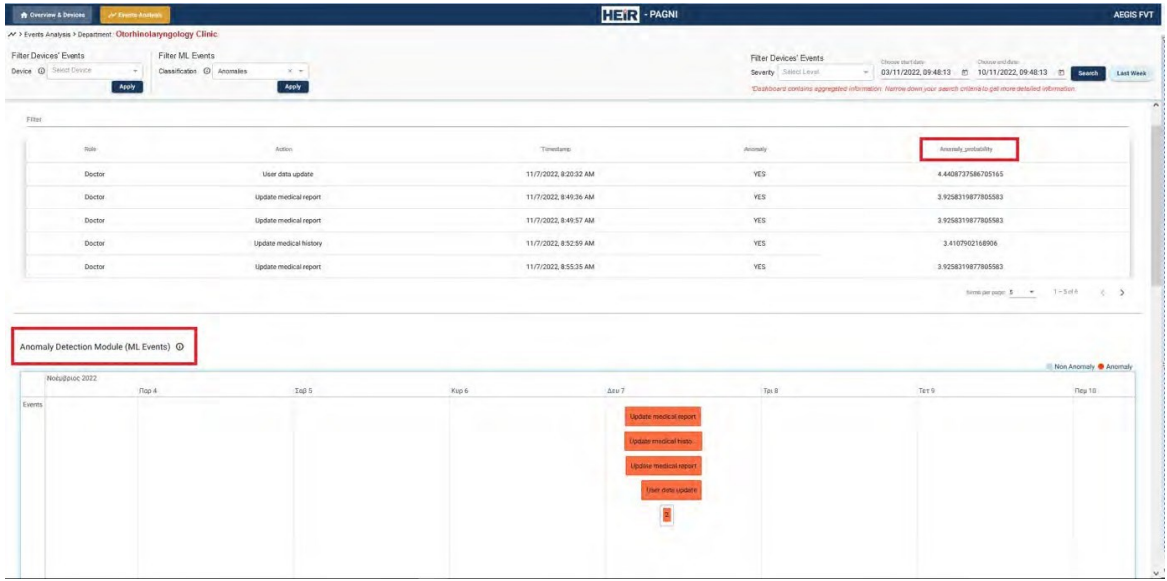

Figure 33: Anomaly detection

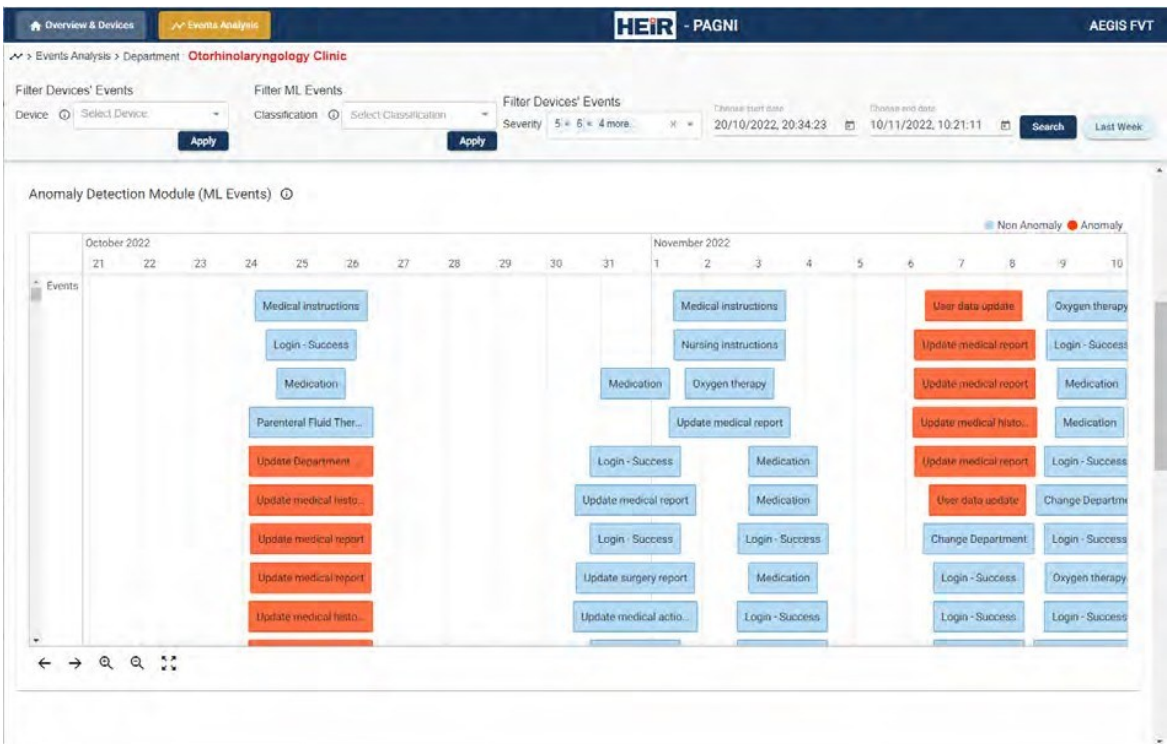

Figure 34: Anomaly detection, filtering

For the specific scenario the HEIR technical team trained a machine learning model with the retrospective logs. The pre-trained model is linked to the real time logs and monitors the logs coming from HIS. When a non-expected behavior has been identified in the real-time logs of HIS from the ML module the 1st Layer GUI of HEIR provides an alert about the HIS logs. The administrator gets a notification for the suspicious behavior from the log files of HIS and opens the 1st Layer GUI of HEIR and at the devices overview, selects the HIS server. Then, checks the temporal representation and the details representation as shown in Figure 33. To get more details, the administrator navigates to the events analysis part and uses the UI of the HEIR platform to filter and view selected time ranges Figure 34.

If the system's activity looks abnormal and other conditions like Memory and CPU usage for the server have been increased, the support team of HIS is informed about the incident for further actions if needed.

### 4.2.2 HYGEIA

HYGEIA aimed to demonstrate the use of the HEIR Framework and specifically of the HEIR Cryptographic Checker (HCC) module, for the proactive managing of system vulnerabilities that can compromise the security and data privacy of its IT infrastructure, especially of the components supporting the "*my-hygeia*" mobile application, that the patients use to access, manage, track, and share data, contained in their Personal Health Records (PHRs). The demonstration took place in the test environment that accurately replicates the sub- section of the IT infrastructure responsible for the operation of the "*my-hygeia*" application. For compliance purposes, the data therein was pseudonymized data conform to regulatory requirements.

#### 4.2.2.1 Infrastructure & Architecture

The PHR application architecture consists of the components displayed in Figure 35 and is accessed (from the mobile device application) via the API Connect Cloud (IBM) and Secure Gateway Client (IBM).

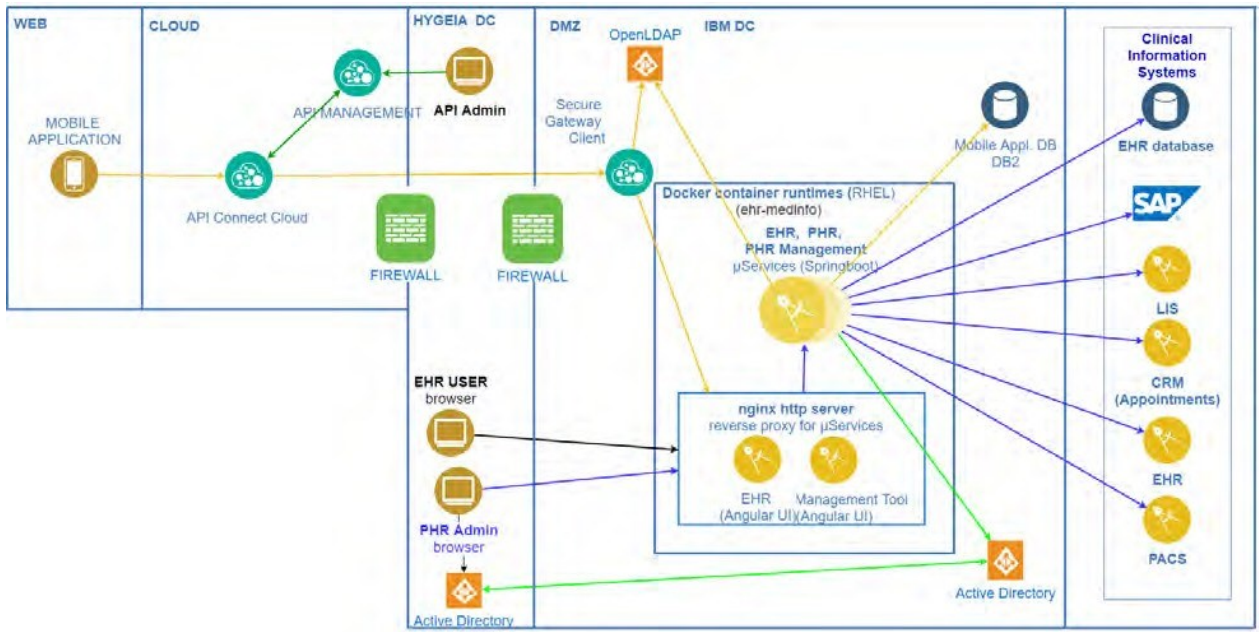

*Figure 35: Hygeia Hospital Use Case Architecture – Logical Diagram*

The following HEIR components have been installed to support the HYGEIA use case: **SIEM, Elastic Connector, Message Broker, SIEM Agents, Forensics Visualization Toolkit (FVT), Visualization Level 1 – (Local RAMA), Dynamic Vulnerability & Monitoring, HNM (Network Module), HET**

(Exploit Tester), HCC (Cryptographic Checker), HEIR client, HEIR Agent, HEIR Aggregator, Local RAMA Score Calculator, ML

### 4.2.2.2 Playbook Scenario

The following section explains the scenario foreseen in the playbook, namely the Cryptographic protocol issue detection using HCC and the RAMA score.

4.2.2.2.1 Cryptographic protocol issue detection using the HCC Module - Involved components: HEIR's Vulnerability Assessment module, HEIR's 1st Layer GUI, HEIR's Local RAMA, HEIR's Cryptographic Checker

The key objective of the HYGEIA use case is for the key actor involved, namely, the PHR Backend System Administrator, to timely detect any devices or servers that are susceptible to cryptographic attacks, and use the information provided by the HCC to resolve the vulnerability.

```json
{
    "ssltest": {
        "description": "",
        "host": "10.18.25.33",
        "sniname": "10.18.25.33",
        "port": "443",
        "protocol": [
            {
                "type": "tls",
                "version": "1.0",
                "enabled": "1"
            },
            {
                "type": "tls",
                "version": "1.1",
                "enabled": "1"
            },
            {
                "type": "tls",
                "version": "1.2",
                "enabled": "1"
            }
        ],
        "heartbleed": [
            {
                "sslversion": "TLSv1.1",
                "vulnerable": "1"
            },
            {
                "sslversion": "TLSv1.0",
                "vulnerable": "1"
            }
        ]
    }
}
```

To this end, the use case demonstrated as follows:

- The Backend System Administrator opens the 1st Layer GUI of HEIR and checks the local RAMA score for any significant reduction.

- In such an occasion, the Administrator detects – through the 1st Layer GUI – the client reporting a vulnerability and checks its details. The Administrator checks the HCC reports (Figure 36) for the port and address of the vulnerable application, and which protocols are affected by the *Heartbleed* vulnerability.

- The Administrator updates the OpenSSL library or the application opening the vulnerable port, depending on which one exposed the vulnerability. This is done outside HEIR environment.

- The Administrator rechecks the local RAMA score, waiting for a new HCC scan to take place, and confirms that the issue has been resolved and no other vulnerability is reported.

*Figure 36: HCC report (JSON format), detecting the Heartbleed vulnerability*

### 4.2.3 NSE and NOKLUS

NSE/NOKLUS aimed to demonstrate the function of the so-called Privacy-Aware Framework, which focuses on facilitating policy-driven access and redaction of healthcare data. The work is based on the HL7 FHIR [32] standard which dictates the rules for digital exchange of medical data. It demonstrated how the PAF can work with data stored either in a FHIR server, or in FHIR JSON format in a PostgreSQL database.

For the actual demonstrations, the following four cases were presented:

- The transfer of data with policy-dictated data transformation to an S3 object store bucket owned by the Norwegian Diabetes Register for Adults (located at NOKLUS).

- How the PAF can work in conjunction with the FHIR Consent resource to enforce time-limited patient consent to data sharing.

- How the PAF can gather data from distributed data registries, to allow seamless, privacy-regulated control to the federated FHIR resources.

- How the PAF can be used to provide role-based access to the metadata stored in the blockchain.

In addition, a professional video [155] was produced as part of the project to provide a basic overview of this use case.

#### *4.2.3.1 Data gathering and collection - Sensotrend Uploader*

The demonstration starts by depicting a patient in her home, plugging in the Continuous glucose monitoring (CGM) transmitter into the laptop (see Figure 37). The video then switches into screen recording mode.

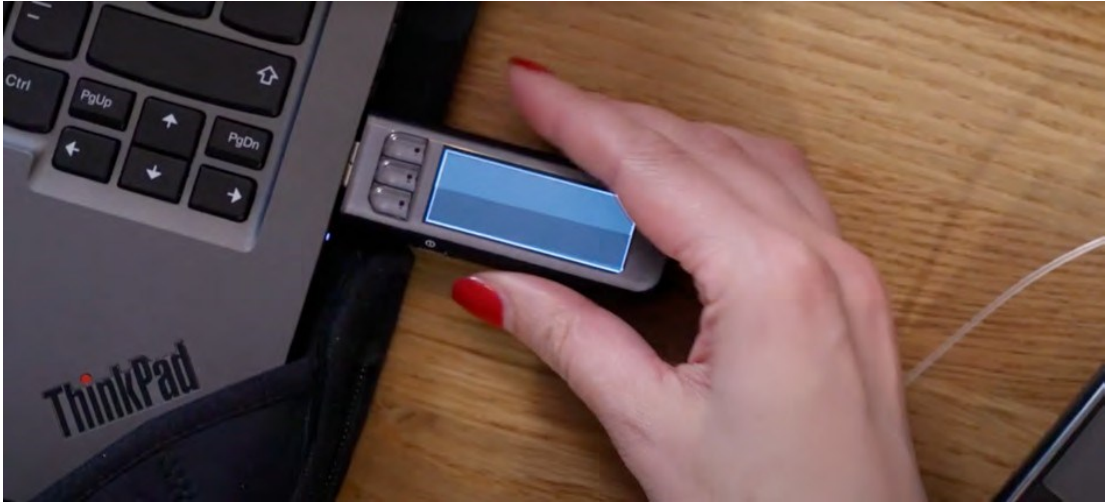

*Figure 37: Sensotrend Uploader – Patient connect CGM transmitter.*

### 4.2.3.2 Data transfer - Sensotrend Uploader, Sensotrend Dashboard

The following screen shows the Sensotrend uploader, where the patient uses her credentials to log in. The upcoming front end displays a selection of devices that can be used with the Sensotrend Uploader and requires the user to pick the device she owns. Subsequently, she triggers uploading all the data gathered (see Figure 38).

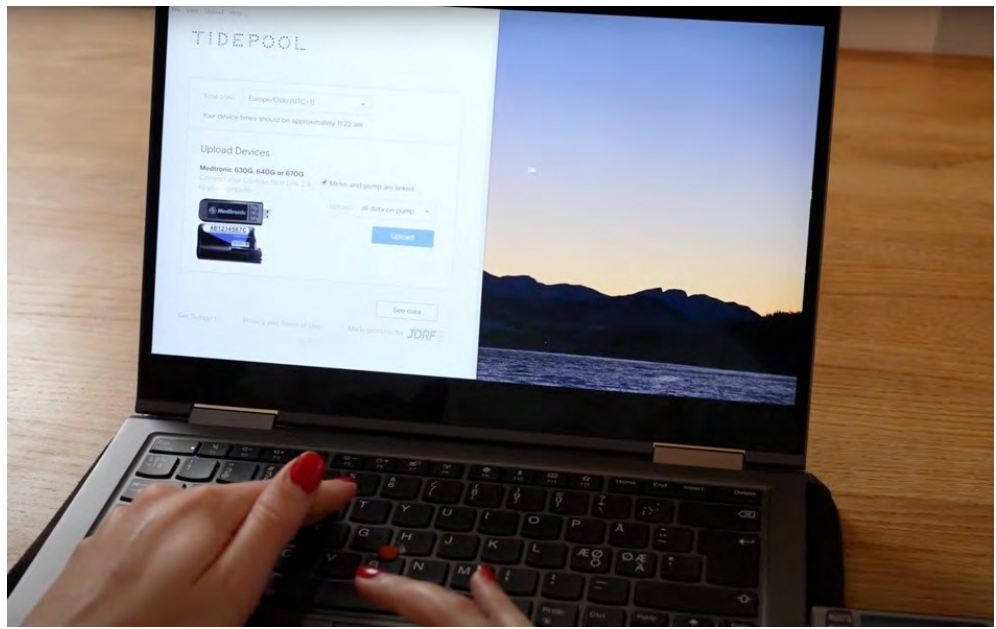

*Figure 38: Sensotrend Uploader – Patient upload CGM device via the laptop*

Once the upload is completed, the patient opens a web browser (this happens in screen recording mode) and logs into the Sensotrend dashboard. The collected and uploaded data is visible in the Dashboard with updated statistics such as – for example – the Glucose Ranges, and the average Glucose and Glucose Variability.

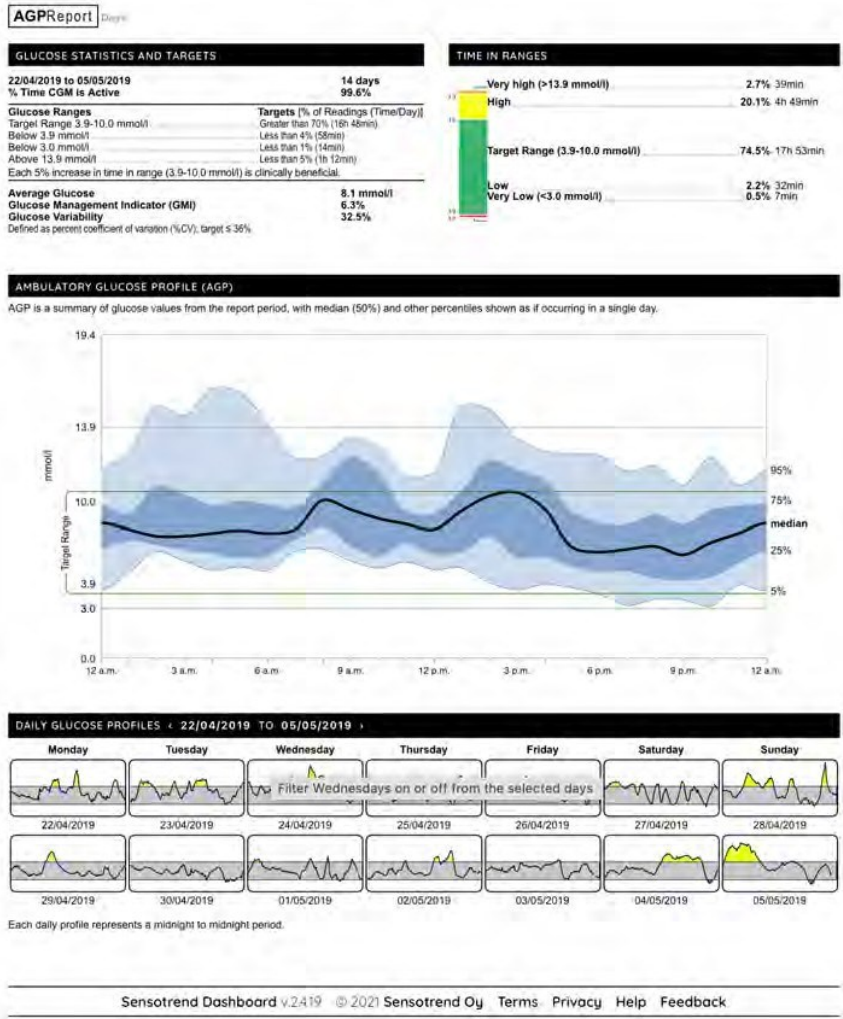

*Figure 39: Sensotrend Dashboard*

The demonstration foresees a brief presentation of the Sensotrend Dashboard (see Figure 39).

### 4.2.3.3 Infrastructure and Architecture

From here on, the demonstration switches from the Patients´ to the Developer's perspective in Microsoft Azure, where the demonstrators guide through the Cloud User Interface (UI) that shows individual servers like DIPS Communicator (see Figure 40), Kubernetes nodes (see Figure 41) and HEIR components (see Figure 42) as well as how they are being set up and the communication between them.

*Figure 40: Azure Infrastructure – DIPS Communicator*

*Figure 41: Azure Infrastructure – PAF and FHIR Sever in Kubernetes*

*Figure 42: Azure Infrastructure – HEIR components*

The first two virtual machines host the PAF and the FHIR-Server. Fast Healthcare Interoperability Resources is a standard that defines the way healthcare information can be exchanged between different computer systems. The third machine runs the DIPS-Communicator, a software that communicates via health- network and transfers aggregated data in this use-case to NOKLUS. The fourth VM contains the different HEIR components such as – for example – the Threat Hunting Module, the Heir Client, or the Local RAMA Score Calculator.

### 4.2.3.4 Playbook Scenarios

The following section explains the four scenarios foreseen in the playbook, namely the Policy- Dictated Data Transformation to an S3 object Store Bucket, Consent management through the PAF´s privacy policy, Distributed data registries using PAF and blockchain technology and Auditing with the PAF using blockchain technology. Within the scenarios, the HEIR Observatory queries open policy agent (OPA) directly for all of the loaded policies (rules) and presents those.

4.2.3.4.1   Policy-dictated data transformation to an S3 object store bucket – Involved components: PAF, HEIR's Observatory

The first use case demonstrated how data from a FHIR-compatible end device (e.g. CGM) is sent to the FHIR server, and from there, the PAF transparently then exports the data with a transformation (statistical analysis) to an S3 store that NOKLUS has access to.

The demonstrators logged in to the MinIO S3, explored the MinIO S3 content and identified the newly redacted data (XML file). In a second step, the XML file that has just been computed based on newly processed data was downloaded and further verified regarding the content inside of the file (e.g., mean, std, and time in range (TIR) glucose values). The file was then picked up by the DIPS-Communicator and transferred to NOKLUS. In case the file could not be verified, the transfer did not proceed, and the process was stopped. The Policy that dictates the data transformation can be easily configured. In the demonstration an example of a policy defined in the so called REGO-language that defined a statistical analysis on top of blood glucose measurements and calculates values such as Time-In-Range, Mean, Standard Deviation etc. was also displayed.

4.2.3.4.2   Consent management through the PAF`s privacy policy - Involved components: PAF, HEIR's Observatory, Sensotrend Dashboard

The second use-case was about the permission to access patient data for research purposes, and was demonstrated through different entities such as hospitals or universities which required access to the patient's data within a certain time frame.

To demonstrate the PAF implementation, the User interface (UI) as depicted in Figure 43 was prepared. This UI lists the consents to share non-anonymized data, that have been previously granted by patients, and additionally allows patients to manage these consents.

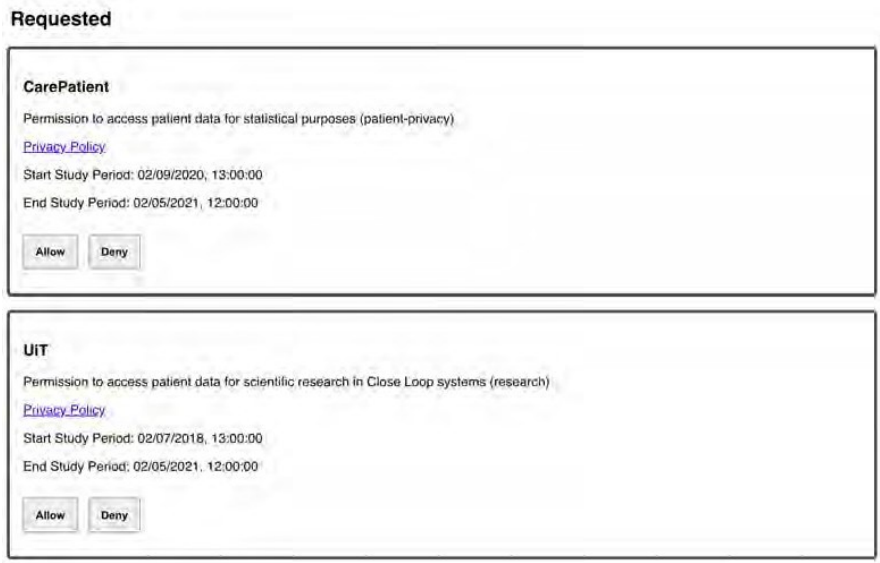

*Figure 43: Consent Management interface implemented in a Sensotrend Dashboard*

The patient now can grant or deny access to his data per requestor such as for example the UiT or NOKLUS. In case the patient permitted access, the requesting entity was provided with the data requested within the specified timeframe.

*Figure 44: PAF - Policy Definition*

*Figure 45: Consenting mechanism – data shared not approved*

Once the PAF received a request for medical observation records from a researcher, it matched the consent with the patients' data and checked whether the date of the data collection falls within the consent timeframe. If this was the case, then the requested data were returned to the researcher. The PAF policy can easily be configured to anonymize personally identifiable information (PII) before it is returned to the researcher as depicted on Figure 45, where subject reference and observation values are transformed into 'XXXX's.

4.2.3.4.3  Distributed data registries using PAF and blockchain technology – Involved components: PAF, HEIR's Observatory

The final case depicts the use of blockchain technologies for the logging of data access actions and demonstrates how the PAF can provide role-based accessing of information.

There were two types of data requesters in this scenario, patients who should be given a collection of all their medical records from the linked registries and researchers, who need access to a large collection of medical records from the registries, but based on policy restrictions, can only access information with PII removed.

For the first type of request, a patient submitted a request for all her medical records from the linked registries. The patient's identity was authenticated by a login process, and a cryptographic certificate was generated and attached to every subsequent request. The certificate contained information about the type of request and embedded the requester's patient id. When the PAF receives the request, it uses securely stored passwords to check all linked registries and retrieve all the records associated with this patient. All data is returned to the patient without filtering or redaction.

For the second type, a researcher issued a request for all records across the registries for a medical study. Once again, the request was accompanied by a cryptographic certificate identifying the researcher. The PAF gathered the medical records, and based on the data categorization previously discussed, replaced PII fields with X's.

#### 4.2.3.4.4 Auditing with the PAF using blockchain technology -Involved components: PAF, Blockchain ledger

For accountability purposes, the PAF logs information about received data requests – who requested the information, when it was requested, the actions performed on the returned data (for example, anonymization) and other parameters to a blockchain ledger. While the blockchain gives a cryptographic guarantee that stored data has not been tampered with, this logged information itself can be considered as sensitive and different roles of users should be given access to different parts of this database. Such as for example, the IT Administrator who is investigating suspicious activity might be given access to all blockchain data, whereas another role may only be allowed to get a summary of data access requests, without any details about the requester or decision taken on the request. A sample output from the Blockchain is used to demonstrate how the log is structured and what kind of information is stored within the log. The sample log contains information about who requested the information, the IP of the requester, the intent of the request, whether the request was authorized or not etc.

```
curl -X GET http://heirauditclient.heirauditingmechanism:8081/queryAllLogs
[{"timestamp":"2022-10-19
15:05:50","userID":"EliotSalant","clientIP":"127.0.0.1","query":"Observation","intent":"research","outcome":"AUTHORIZED","policyDec
ision":"[{'action': 'JoinResource', 'description': 'Perform a JOIN', 'joinTable': 'Consent', 'whereclause': ' WHERE (
(observation.issued BETWEEN consent.provision_provision_0_period_start AND consent.provision_provision_0_period_end) AND
consent.status='active' ) OR ( consent.status='draft' OR consent.status= 'proposed' OR consent.status='rejected' OR
consent.status='inactive' OR consent.status='entered-in-error' ) ', 'joinStatement': ' JOIN consent ON
observation.subject_reference = consent.patient_reference '}, {'action': 'RedactColumn', 'description': 'redact columns:
[valueQuantity.value subject.reference]', 'intent': 'research', 'columns': ['valueQuantity.value', 'subject.reference'], 'options':
{'redactValue':
'XXXXX'}}]","encryptedLog":"ndVnHLiaiPCjBiDoiZO0/yfqQxmZ2UAgNsg/S+wcyDaL9uSjjqNY1PMmTHU07mgl3XLruSMPfGxILGLIO58Q778BEsqs+Dsg3+y0GXm
+ARXcHrcgz5LUDVVH3JEMUeLz5Np0Ls4YxMfSZLIKp2FjpdCXZRQYSFnaNdcx0PoPdDwty7+f4WKtkAGV1IDWup24RF8UNmlhoqgpQtn10uE+JR0GoB4lLuiFCCTNCivOgn
8J505t0xPYIt0SUgbJuUYF357ZLEo6AWdJRuFs/Z1w70CD/iGp5T0RXYOdtirBEu5cbJRimkBAEM2QkwQzxTZXnpaQVx4d8oALSd6vmSV6dprxnUpgEpQTrhLiWKYE0p4ya
I8F8TK1M8XPREDOa1Bb+Q7FKWhuXNKwgfyXik17H9HwZP/Di1sF4OYB/wzEAcI66RqhPrqpGPY2ZNDtvpI9iA9NOZY8807u5uxmK6IKbJYXr+ielnZ5cnnSG3SbQrXCBo1v
xD914vsPoZlAQFKU3THgRfY1BNsBVltITafn4GjWh4nRC4o/2CM2xFty68WZpmJjei2Kd/GkcLH1WHiIypZbrcxp8026Aoi7F+pX45DcodQ8n21dPjDF2TRldk2m56JlOfz
LH+yBNkWQMKdyH3ISfjGj3zPgbvIH3ZAQgIBTAK5UPTzAicByXL1SvN2NqtouxS51gMmykZmyB+KuMre+e0Tt/tFYTTz84YqOYziMT9M/2vZmorP9VRXYWld+IC/wRMxtre
Y/+pH9Y6E8UeO5LeiFKPpjsVbCsDcMekLf6dtp2jN3xT9pB3OU02/e4S0oso9xFruWxGUf2Sg6kK+m9Cq+g9IE8Mus1mK7AjfBii6TDThV7joA1/J9ochlS5MYKb33h15lo
7XbP2AZ3SfW3oT47oBbUA59TbPB5eUW8dfAbE89SeztmFyiYnNVrakZJM0yDx63hKrNfcWTCIB1eh4Mg9bFWAQARTzEtzGjsce7T1Z0bq5LhTdkqwGLCvdkPNNJ3chGdeUz
G2zEVCxEDZefMacN7Ao/LfW/vfJ3uyTZ4SWZDUUxEhMYQFq5rWttkhiB1RR18piN0rAXczyAIIaNZ8g7m7D/L4e8b9Ia1lDIWKvF4COxduFjazUuF3VgBv/ejtjRCD88FpE
sqjp/lHQDYgjt9twV9VM6zfwCjPF4P2s0sjI+YIyqaY3a9QEmiXDKmh5GvdoC1DHPcTJmR/hA7AMnBwV6C8z4ng=="}]
```

*Figure 46: Blockchain entry*

### 4.2.4 CUH

Croydon University Hospital aimed at demonstrating the functionality of specific HEIR components that monitor data flows from medical devices inside systems that exist inside a medical facility. The main focus was the detection of aberrant data flows indicative of a malfunctioning medical device that was intended to represent a medical device compromised by malware. We tested HEIR agent's ability to detect abnormal signals that may arise from an intrusion of such malware-affected devices, in this case the aberrant signals arising from the TEAM 3 [154] device. This device is used to monitor maternal and fetal heart rate, maternal oxygenation saturation, as well as uterine activity (CTG). By reading these values, a doctor is able to monitor the well-being of mother and baby during the pregnancy and labour, leading to timely intervention if needed to ensure safe delivery of the baby or preservation of maternal health. The scenario for this playbook is that the TEAM 3 device has been compromised, thereby generating aberrant signals which could then compromise patient safety. It is therefore anticipated that HEIR would be able to capture these aberrant signals, reflective of a compromised medical device, and help reassure end users that compromised medical devices may be detected when the HEIR system is in use.

### 4.2.4.1  Infrastructure and Architecture

The following is the architecture that would be used for testing the HEIR system on site.

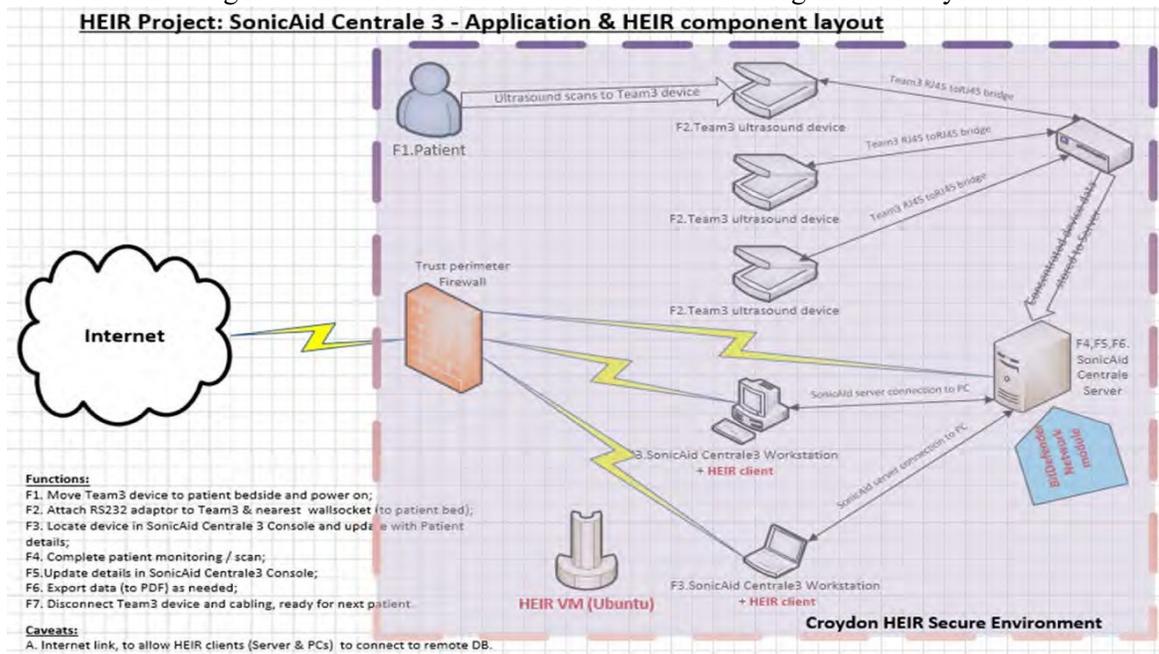

Figure 47: CUH Infrastructure architecture

### 4.2.4.2  Playbook Scenario

The following section explains the scenario foreseen in the playbook, namely the Team 3 device infection and Threat detection through HEIR´s Machine Learning.

4.2.4.2.1  Team 3 device infection and Threat detection through HEIR´s Machine Learning - Involved components: Team 3 device, HEIR´s Machine Learning

For the actual demonstration, we initially demonstrated a TEAM 3 device in normal operation, and the HEIR user interface highlighting normal signals being detected. We would then simulate an abnormal device output, to identify the changes seen within the user interface (reflecting the abnormal signals). Since an anomaly was detected, actions are needed from the clinical team to check on the device that is being monitored. Finally, we would revert back to normal TEAM 3 output, which would result in normal signals being depicted again in the user interface.

The demonstration started with the training of our machine learning algorithm. A fully anonymized database with historical data was built, from a real TEAM 3 device operating inside the Croydon University labour ward for over 6 months. For this process, all data were obtained after data control approval. This data extract of the defined biophysiological components (heart rate, uterine contractions, pulse oximetry) was used to train the system, along with synthetically generated data of abnormal profiles, so that detection of anomalies could be established that could be distinguished from normal profiles. These trained components were then installed into HEIR pilot deployment, where an isolated TEAM 3 device and a mock infrastructure reflective of current working IT systems, was implemented.

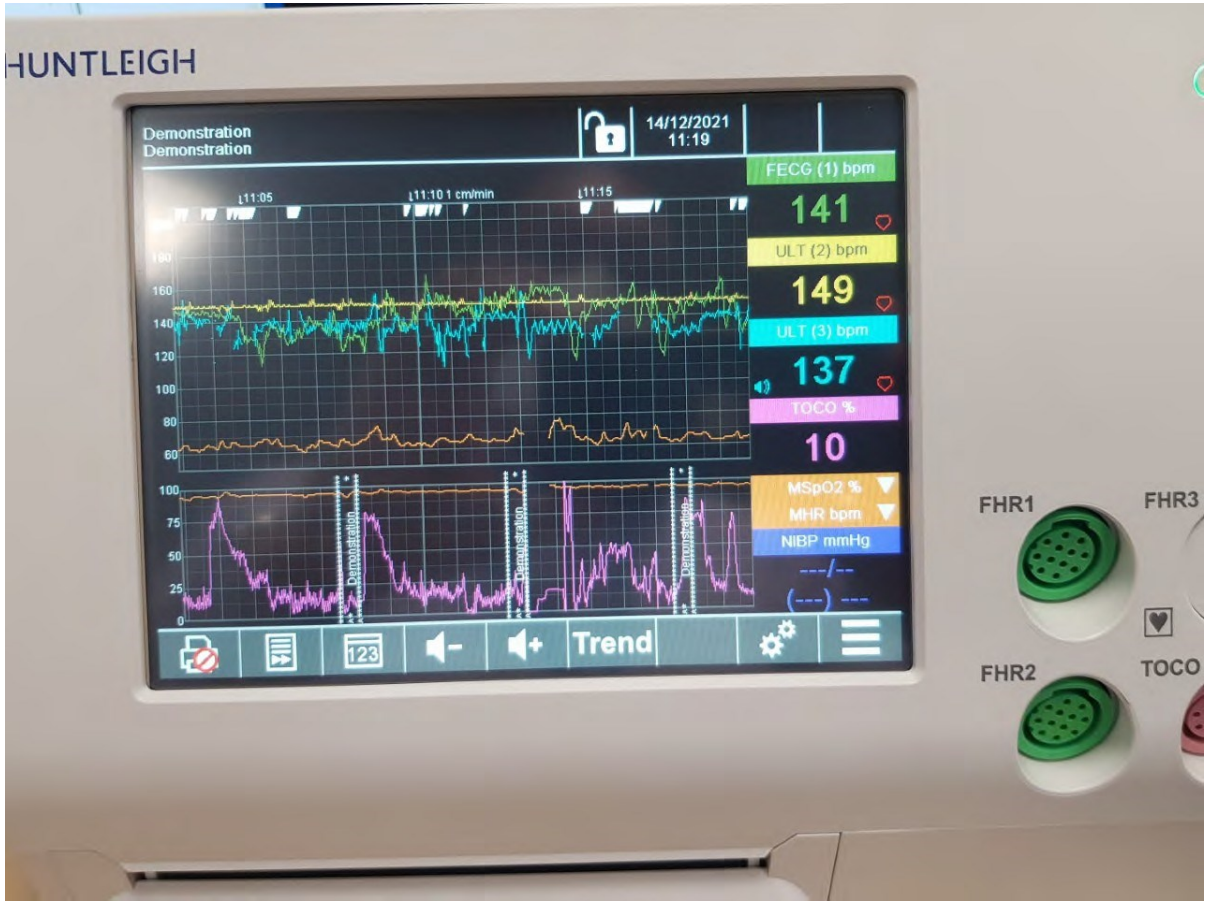

*Figure 48: Team3 device in operation*

# 5 Conclusions

HEIR is trying to tackle one of the most important issues of today's world: cybersecurity in the healthcare domain. To enhance the security of healthcare environments and mitigate risks, there is a demand for a comprehensive end-to-end cybersecurity approach. HEIR offers a holistic cyber-intelligence platform for healthcare environments, facilitating secure data exchange across healthcare and research organizations with high levels of resilience, reliability, accountability, and trustworthiness. The platform addresses threat prevention, detection, mitigation, and real-time response.

In this paper we provided a comprehensive outline of the integrated platform, covering various aspects such as the system's architecture, the HEIR modules integrated within it, the communication between these modules, and how they support use case scenarios and playbooks. Additionally, the paper also describes details on the methodology employed, system and network specifications, security risks and measures taken to mitigate them, as well as the results of technical testing.

Considering the system's market adoption, the primary value proposition of HEIR stems from its extensive range of advanced offerings. Firstly, it provides real-time threat-hunting services, protecting healthcare organizations against the ever-increasing threat landscape. Secondly, HEIR enables improved and secure data-sharing, through its privacy-aware framework. This framework strongly emphasizes the trustworthiness of sensitive data and facilitates secure and confidential sharing of such data. It ensures that privacy and data protection standards are effectively upheld, providing stakeholders with confidence in handling sensitive information. Lastly, HEIR provides the Observatory for the Security of Electronic Medical Devices. This observatory serves as an intelligent knowledge base accessible to a wide range of stakeholders.

This presents an opportune moment for HEIR to capitalize on key challenges and market drivers, adopting a customer-centric value approach and establishing capabilities to support business modelling for each market. Despite tough competition, HEIR distinguishes itself technologically as a platform supporting multi-level security in healthcare environments, encompassing different technologies to build the Risk Assessment for Medical Applications (RAMA). It boasts multiple novel components, employing highly innovative approaches to address several security requirements specific to industrial environments facing the complex challenges of the healthcare domain.

HEIR also holds a unique business advantage compared to existing solutions. It offers a granular approach and caters to various security requirements at all levels, enabling potential users to selectively choose and pay only for the novel components they need from a vast list of HEIR components. This flexibility allows for tailored solutions that precisely address the specific needs of involved industrial sites.